# Determining sensor geometry and gain in a wearable MEG system

Ryan M. Hill[1,2,+], Gonzalo Reina Rivero[1,2,+], Ashley J. Tyler[2,+], Holly Schofield[1,2,+], Cody Doyle[3], James Osborne[3], David Bobela[3], Lukas Rier[1,2], Joseph Gibson[1], Zoe Tanner[1,2], Elena Boto[1,2], Richard Bowtell[1,2], Matthew J. Brookes[1,2], Vishal Shah[3,+], and Niall Holmes[1,2,+]*

[1]Sir Peter Mansfield Imaging Centre, School of Physics and Astronomy, University of Nottingham, University Park, Nottingham, NG7 2RD, UK
[2]Cerca Magnetics Limited, Units 7&8 Castlebridge Office Village, Kirtley Drive, Nottingham, NG7 1LD, Nottingham, UK
[3]QuSpin Inc. 331 South 104th Street, Suite 130, Louisville, Colorado, 80027, USA

+ Indicates equal contribution from multiple authors

* Indicates correspondence to:

    Dr. Niall Holmes,

    Sir Peter Mansfield Imaging Centre,

    School of Physics and Astronomy,

    University of Nottingham,

    University Park,

    Nottingham NG7 2RD,

    United Kingdom

    E-mail: niall.holmes@nottingham.ac.uk

**ABSTRACT**

Optically-pumped magnetometers (OPMs) are compact and lightweight sensors that can measure magnetic fields generated by current flow in neuronal assemblies in the brain. Such sensors enable construction of magnetoencephalography (MEG) instrumentation, with significant advantages over conventional MEG devices including adaptability to head size, enhanced movement tolerance, lower complexity and improved data quality. However, realising the potential of OPMs depends on our ability to perform system calibration – which means finding sensor locations, orientations, and the relationship between the sensor output and magnetic field (termed sensor gain). Such calibration is complex in OPM-MEG since, for example, OPM placement can change from subject to subject (unlike in conventional MEG where sensor locations/orientations are fixed). Here, we present two methods for calibration, both based on generating well-characterised magnetic fields across a sensor array. Our first device (the HALO) is a head mounted system that generates dipole-like fields from a set of coils. Our second (the matrix coil (MC)) generates fields using coils embedded in the walls of a magnetically shielded room. Our results show that both methods offer an accurate means to calibrate an OPM array (e.g. sensor locations within 2-mm of the ground truth) and that the calibrations produced by the two methods agree strongly with each other. When applied to data from human MEG experiments, both methods offer improved signal-to-noise ratio after beamforming suggesting that they give calibration parameters closer to the ground truth than factory settings and presumed physical sensor coordinates and orientations. Both techniques are practical and easy to integrate into real-world MEG applications. This advances the field significantly closer to the routine use of OPMs for MEG recording.

# 1) INTRODUCTION

Optically-Pumped Magnetometers (OPMs) have emerged as a useful means to measure magnetic fields generated by current flow in neural assemblies in the brain (see Brookes et al. (2022) for a review). These assessments of brain activity – termed magnetoencephalography (MEG) – have traditionally been made using sensors based on Superconducting Quantum Interference Devices (SQUIDs) (Baillet, 2017; Hämäläinen et al., 1993). However, SQUIDs operate at cryogenic temperatures, leading to several limitations. First, sensors must be fixed in position in an array that cannot be adapted to individual head size and results in a greater brain-to-sensor distance (and lower signal) for smaller heads (e.g., infants). Even in subjects who fit the array (i.e., adults), sensors must be distal to the head to accommodate thermal insulation, again limiting sensitivity. Second, movement of the head relative to the fixed array degrades the quality of data, meaning subjects must remain still for long periods. Finally, cryogenic systems are complex, requiring either regular filling with liquid helium and/or a helium reliquefier. In contrast, OPMs exploit the quantum properties of alkali atoms to measure the neuromagnetic field (see Schofield et al. (2023) and Tierney et al. (2019) for reviews). They do not require cryogenics and can be microfabricated into small (sugar cube-sized) packages (Schwindt et al., 2007, 2004; Shah et al., 2007; Shah and Wakai, 2013). This offers the possibility of a simpler sensor array that is lightweight, adapts to head shape/size (Corvilain et al., 2024; Feys et al., 2022; Hill et al., 2019; Rier et al., 2024), and enables movement during a scan (Boto et al., 2018; Holmes et al., 2023b; Mellor et al., 2023). Because the sensors don't require low temperatures, they can be sited closer to the head, ostensibly improving sensitivity and spatial precision (Hill et al., 2024; Iivanainen et al., 2017). These advantages suggest that OPMs could overtake SQUIDs as the fundamental building block of MEG instrumentation.

An OPM contains a glass cell housing a vapour of alkali atoms. Laser light is passed through the cell at a wavelength resonant with the D1 energy transition. This, coupled with circular polarisation of the laser light, enables atoms to be "pumped" into single quantum state, such that their atomic magnetic moments align (Happer, 1972). In this way, the vapour gains a bulk magnetisation, effectively becoming "magnetic", with coherence (across atoms) maintained via operation of the system in a spin exchange relaxation free (SERF) regime (Allred et al., 2002). Assuming the sensor is in zero magnetic field, the atoms remain trapped in their single state and can no longer absorb photons. This means the transmission of light through the cell is maximised. However, if there is an interaction between the bulk magnetisation and an external magnetic field, then the atoms change state and can once again absorb photons, causing a change in light transmission through the cell. Specifically, the polarisation of the vapour (which relates to the opacity of the cell) can be described by the Bloch equations. In this simple case, the intensity of the light passing through the cell becomes a Lorentzian function of magnetic field. When a sinusoidal 'modulation' field is also applied to the cell in a direction perpendicular to the laser beam (e.g., in the x-direction) at a frequency of around 1 kHz, the solution to the Bloch equations is modified such that the intensity of the light passing through the cell oscillates at the modulation frequency, with a modulation amplitude that is a linear function of the ambient x-oriented magnetic field, for a range of fields close to zero (~±1 nT) (Cohen-Tannoudji et al., 1970). When the modulation of the transmitted light intensity is measured (e.g., by lock-in detection at the modulation frequency) the system becomes a magnetometer with directional sensitivity. The use of two sinusoidally varying fields oriented at right angles to one another (both perpendicular to the laser beam) and applied in quadrature at the same frequency, allows two Cartesian components of the vector field (e.g., x and y) to be measured simultaneously. The field component along the direction of the laser cannot be measured with equivalent accuracy. However, the use of two orthogonal laser beams allows simultaneous measurement of all three field components (Boto et al., 2022; Shah et al., 2020). OPMs are sensitive to small fields; with sensitivities ~10-25 fT/√Hz (this is higher than the ~3-5 fT/√Hz commonly achieved using SQUIDs but is typically compensated by increased signal due to the closer proximity of sensors to the brain).

Despite their promise, OPMs still have fundamental challenges to overcome prior to becoming mainstream MEG technology. One key challenge is array calibration, which for MEG applications includes defining the location of sensors (relative to each other), the precise orientation of the sensitive axis, and the relationship between the (voltage) output of the sensor and the magnetic field that passes through the cell (the sensor gain). All three must be known with high accuracy for MEG signals to be accurately recorded. In a SQUID-based MEG system, these calibration coefficients are set during system construction/installation, typically via the use of known magnetic fields from electromagnetic coils placed either inside or outside (Li et al., 2016) the system. Calibration accuracy depends on, for example, the tolerance of superconducting windings and their (fixed) placement within a cryogenic dewar, as well as deviations between the mathematical model of the field produced by the calibration coil and the actual field produced by the coil. Techniques such as harmonic model-refinements (Chella et al., 2012; Taulu et al., 2005), taking X-Ray CT scans of coil wire paths (Oyama et al., 2022), and using larger coils to reduce field model errors (Adachi et al., 2023) can be used to improve accuracy.

In OPMs, however, calibration is more challenging. At the simplest levels, this is because the relative position and orientations of the sensors can vary from subject to subject. For example, when switching sensors between different rigid helmets. However, calibration is further complicated by the physics of OPMs, which can cause changes in both gain and orientation sensitivity. Specifically, calibration of an OPM depends on three key factors: (1) manufacturing tolerances, (2) factory-programmed sensor operational settings, and (3) local operating conditions. **Manufacturing tolerances** include the placement of the on-sensor coils with respect to the vapour cell, the homogeneity and orthogonality of the fields these coils generate, the propagation direction of the laser beam(s), the position where the vapour cell and the laser beam intersect, the laser beam diameter, polarisation, and intensity, and vapour cell characteristics (such as buffer gas pressure). **Operational settings** include the relative phase and frequency of the modulation field produced by the on-sensor coils, the exact lock-in demodulation procedure, the vapour cell temperature, the algorithm used to zero the field within the vapour cell and the laser wavelength offset settings; all of which result in light shifts and other Rubidium density factors that induce asymmetries in the resonance lineshape. **Local** **sensor operating conditions** include temporal field drift and varying background magnetic field and gradient field profiles between sites (Hill et al., 2022) which can introduce cross-axis projection errors (Borna et al., 2022) that strongly impact the orientation of the sensitive axis and the sensor gain. Sensor age and intentional (or unintentional) changes in sensor firmware or settings also have an effect. For these reasons, it is inherently unreliable to always use the physical sensor orientations, or sensor factory gain settings for OPM calibration. Instead, sensor positions, orientations, and gains should ideally be independently determined during each MEG experiment to identify any deviations in sensor behaviour.

In this paper, we take a newly developed 192-channel miniaturised integrated OPM-MEG system (Schofield et al., 2024) and introduce two independent means of calibration – both based on generation of well-characterised magnetic fields over the OPM sensor array. Our first method is a helmet-mounted planar disc containing dipolar coils called the 'HALO'; our second method uses a matrix coil (MC) magnetic field nulling system (Holmes et al., 2023b) (similar to work by Iivanainen and colleagues (Iivanainen et al., 2022) to perform similar calibrations). We aimed to show that we can accurately determine sensor locations, orientations and gains, and in this way calibrate all sensors in an array.

The calibration was tested by setting up the array in a rigid helmet and quantitatively comparing the HALO- and MC-derived sensor locations, orientations, and gains to those obtained from the computer aided design (CAD) model of the helmet and presumed factory settings. The calibration coefficients are tested for robustness using multiple measurements with the same technique. Further, our independent calibration methods are compared against each other. Finally, we undertake a series of MEG experiments in human participants, testing the hypothesis that the accuracy of mathematical modelling of MEG data will increase following calibration (using both calibration systems).

## 2) METHODS

### 2.1) OPM-MEG system

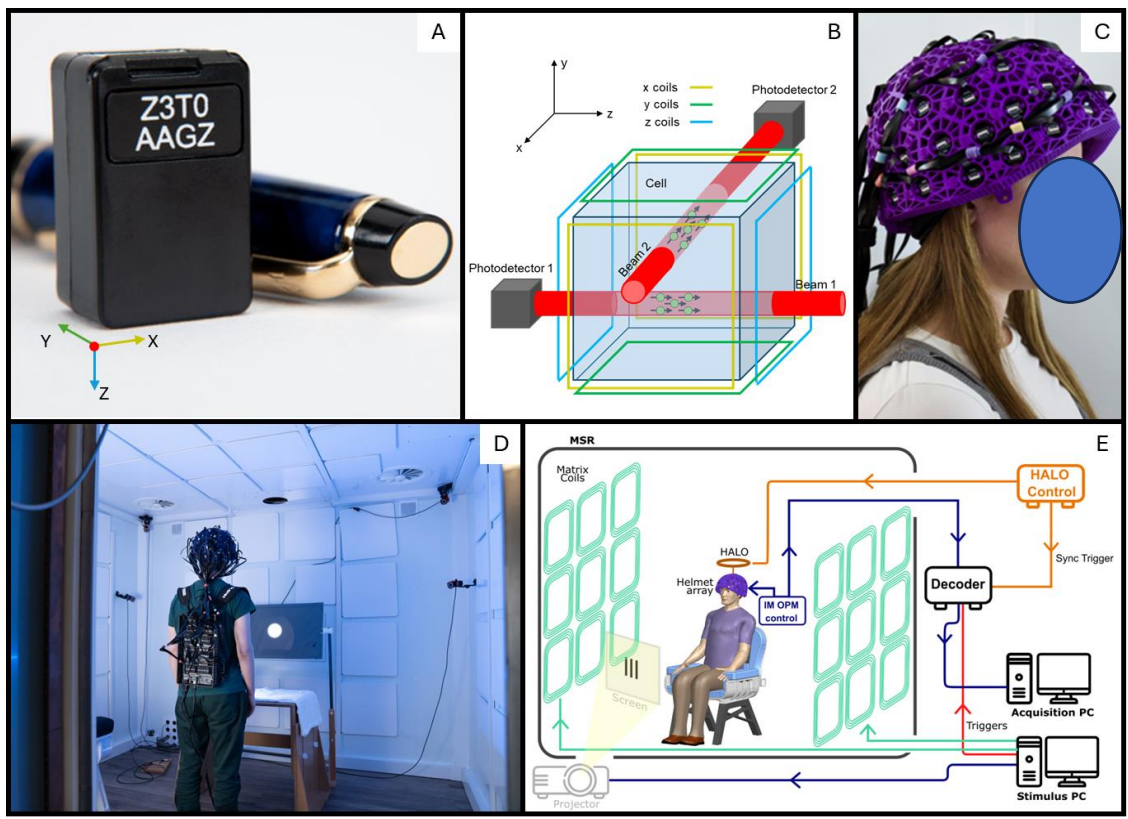

***Figure 1: OPM-MEG system:*** *A) Photograph of a triaxial OPM sensor. B) Schematic diagram showing a triaxial sensor. Two orthogonal laser beams intersect the cell and enable measurement of field in three orientations from two photodiodes. C) 3D printed helmet in which sensors are mounted. D) The integrated miniaturised electronics system formed as a backpack which can be worn by the subject. The white square MC panels can be seen on each face of the MSR. E) System schematic.*

We used an OPM-MEG system with 64 triaxial sensors, which can each measure the magnetic field along three (notionally) orthogonal orientations (QuSpin Inc. Colorado, USA). This allows data to be collected from 192 independent channels. Each sensor head (Figure 1A) incorporates a $^{87}$Rb vapour cell, a laser (tuned to 795 nm – the D1 transition for $^{87}$Rb) for optical pumping, on-board electromagnetic coils for field control (including generation of modulation fields) and two photodetectors for signal readout. The size of the sensor is 12.4 x 16.6 x 24.4 mm$^3$ and each sensor weighs 4 g. Inside the sensor (Figure 1B), a beam splitter and associated optics splits the laser output and projects two orthogonal beams through the cell (Boto et al., 2022; Shah et al., 2020). As noted above, field measurements are most accurate perpendicular to the beam, and so beam 1 allows field measurement in the x- and y-orientations, whilst beam 2 enables measurement in the y- and z-orientations. The two beams combined offer complete assessment of the field vector (for a more complete description, see Boto et al. (2022)). The sensors were mounted in a 3D-printed helmet (Cerca Magnetics Limited, Nottingham, UK) which offers approximately even coverage of the cortex (Figure 1C). The helmet is designed using CAD and constructed using 3D-printing, so the *assumed* locations of the vapour cells in each sensor are known (in principle) to an accuracy of ±1 mm. Sensor orientations are assumed to be perpendicular to the OPM housing (x, y, and z in Figure 1A) and are therefore known (in principle) to ±1°.

Control of the OPM requires circuitry and associated software to stabilise the cell and laser temperatures, identify and lock the laser wavelength to the D1 resonance, zero the magnetic field within the cell, generate modulation fields and read out signals from the photodetectors (which are modulated at the same frequency as the modulation fields, and captured using lock-in-detection). Here, control of these processes was achieved using an integrated miniaturised electronics unit (see (Schofield et al., 2024) (Figure 1D) ("NEURO-1", QuSpin, Colorado, USA). The electronics unit is 36 x 20 x 6 cm in size, weighs 1.8 kg and can be worn as a backpack to enable (for example) ambulatory studies (Figure 1D). Each sensor head was connected to the backpack by a ribbon cable (2.2 gm$^{-1}$ and 90 cm in length) and each sensor is controlled by a separate electronics card. These cards are grouped together in modules of 8. The digitised output of each sensor/module is sent to a multiplexer and then to a network card which passes all data, via ethernet, to a decoder (DAQ – sbRIO9637, National Instruments). The decoder integrates the sensor outputs with peripheral signals (e.g., those carrying the timings of the experimental paradigm) and passes everything to an acquisition PC, via a second ethernet connection. It is noteworthy that the electronics also enables three-axis closed-loop sensor operation, enabling a large dynamic range (currently ±8 nT). However, closed loop operation was not required for the work described here due to low magnetic field drifts (Hill et al., 2022). The bandwidth of the OPMs is 0 – 135 Hz (where the upper limit represents a 3 dB loss in sensitivity).

The sensor array and backpack were housed in a magnetically shielded room (MSR) comprising four MuMetal layers and one copper layer, which attenuate DC/low frequency and high frequency magnetic interference fields respectively (MuRoom, Magnetic Shields Limited, Kent, UK). The MSR walls were equipped with degaussing coils to reduce any remnant magnetisation prior to data collection (Altarev et al., 2015). The MSR was also equipped with the MC (Holmes et al., 2022; Holmes et al., 2023). This comprised coil elements attached to the inner walls of the MSR. These elements can be independently energised using low noise current drivers (Magnetic Shields Limited, Kent, UK) such that the field from all elements sums to flexibly create controlled magnetic field patterns within the MSR (this system will be described in more detail below). A single 'acquisition' computer was used for OPM-MEG data acquisition. Experimental paradigms (along with associated triggers) and the MC were controlled by a second 'stimulus computer'. A schematic diagram of the system is shown in Figure 1E.

**2.2) Calibration with HALO**

The "HALO" comprised a circular printed circuit board containing 16 independently controllable electromagnetic coils. Each coil was formed from 6-layers of a 45-turn spiral (270 turns total), with an inner radius of 3.68 mm and an outer radius of 10.94 mm. When energised (at sufficient distance from a sensor), the coil creates a field that closely resembles a magnetic dipole. (Note that the large number of turns means a small current can be applied, and so the relative contribution of stray field from the cables feeding current to the coil is minimised.) 12 of these dipolar coils were equidistantly spaced (every 30°) around a circle of radius 12.7 cm; a further 3 dipoles were spaced at 120° intervals around a circle of 6.3 cm radius; the final dipole was placed at the centre of the circle (see Figure 2 (left)). The HALO was electrically coupled to a driving circuit (QuSpin Inc.).

The procedure for running the HALO was as follows. First, the HALO was mounted in the OPM helmet on a base which holds the disc 5 cm above the helmet, as shown in Figure 2 (right). Following this, the door to the MSR was closed and the room demagnetised. Each of the 16 dipoles was sequentially energised, to create 16 (spatially different) dipolar field patterns across the sensor array. The driving circuit generated a 10 Hz oscillating current which was applied to one dipole for 1 s before being switched to the next dipole. Once all 16 dipoles had been energised the current was turned off. One limitation of dipoles is that the magnetic field falls very rapidly with distance, and so whilst nearby OPMs see a large field, distal OPMs see a small field. To overcome this, the sequence was run 4 times with dipole moments of 4 µAm$^2$, 11.8 µAm$^2$, 28.3 µAm$^2$ and 41.0 µAm$^2$ with a 5-s break between sequences. The procedure lasted ~90 s and OPM data were collected throughout.

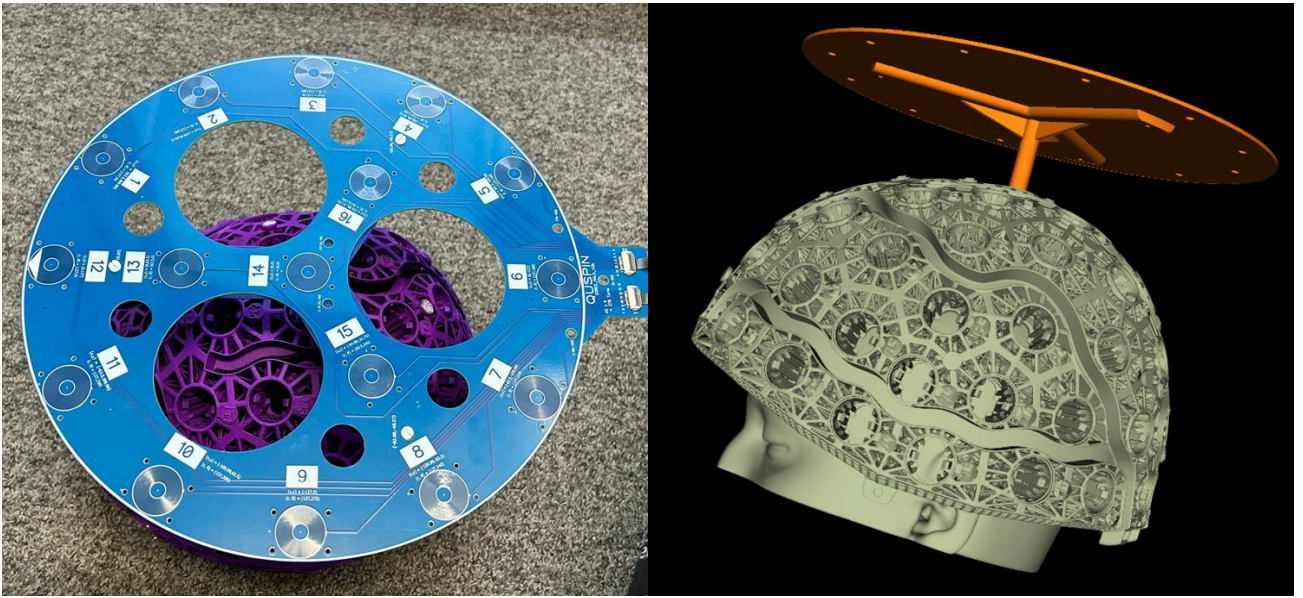

*Figure 2: HALO calibration system: Left) Photo of the HALO. The system contains 16 independently controllable magnetic dipoles which make spatially distinct field patterns across the OPM array. Right) CAD representation of the HALO in position in the OPM helmet.*

An algorithm was generated to calibrate each OPM based on the model of the field generated by the HALO. Assuming dipole $n$, is at location $\boldsymbol{p_n}$ relative to the centre of the HALO and that a sensor is located at $\boldsymbol{r}$ (also relative to the centre of the HALO) and makes a single measurement along an axis defined by the unit vector $\hat{\boldsymbol{o}}$. The sensor measurement, $V_n(\boldsymbol{r},\hat{\boldsymbol{o}})$ (i.e. in volts), can be modelled by the equation

$$V_n(\boldsymbol{r},\hat{\boldsymbol{o}}) = g\left\{\frac{\mu_0}{4\pi}\left[\frac{3(\boldsymbol{r}-\boldsymbol{p})(\boldsymbol{m}.(\boldsymbol{r}-\boldsymbol{p}))}{\delta^5} - \frac{\boldsymbol{m}}{\delta^3}\right]\cdot\hat{\boldsymbol{o}}\right\}. \qquad [1]$$

Here, $\boldsymbol{m}$ is the magnetic dipole moment (whose orientation is perpendicular to the plane of the HALO, and magnitude is assumed to be the same for all dipoles), $\delta = \|\boldsymbol{r}-\boldsymbol{p}\|$ is the Euclidean distance from the coil origin to the measurement location, and $\boldsymbol{r}$. $g$ represents the sensor gain (in units of V/T). This equation allows us to model the fields generated by any dipole in the HALO, at any location and orientation across the array.

A single sensor makes three measurements of field ($[V_{n\hat{o}_1}, V_{n\hat{o}_2}, V_{n\hat{o}_3}]$ which represent the field magnitude along three axes, $\hat{o}_1$, $\hat{o}_2$ and $\hat{o}_3$). By measuring field from all 16 dipoles, a single sensor makes 48 separate field measurements, and this is repeated 4 times (due to the 4 different dipole amplitudes) making a total of 192 measurements per sensor. By fitting these values to a model generated by the repeated use of Eq. [1], it becomes possible to determine the 12 free parameters comprising the sensor location ($r$), the 3 measurement orientations ($\hat{o}_1$, $\hat{o}_2$ and $\hat{o}_3$), and the 3 gains ($g_1$, $g_2$ and $g_3$). (In practice we defined orientation vectors such that their the magnitude represented the gain and the normalised unit vector is the orientation, hence only 12 free parameters are needed not 15). This offers a complete calibration of an OPM.

In practice, for each signal, the amplitudes and direction of the measured signals (relative to the phase of the coil current) were estimated via a fast Fourier transform. To avoid the problem of signals being too small in distal sensors, or so large in proximal sensors that CAPE effects and sensor gain errors would confound the calibration, only segments of data where the measured 10 Hz signal was between 1 pT and 1,000 pT in amplitude were used in the fitting algorithm. The fitting procedure was implemented in MATLAB using the 'fmincon' function to minimise the sum of the squared differences between the model and measured data. The initial guess for the sensor position was the same for each sensor and was approximately the centre of mass of the helmet. The initial channel gain was set to the assumed 2.7 V/nT and each channel orientation was set in the front-back direction of the helmet. [Note – as shown in Figure 1B the two laser beams, and hence the precise locations of the field measurements for the x and z field components are offset from one another by ~600 μm. However, this small difference was ignored in the fitting algorithm.]

**2.3) Calibration with matrix coils**

The MC comprises 94 independently controllable square coils of side length 450 mm arranged (approximately) symmetrically in 4x4 grids on each MSR face. (Due to access requirements, coils on the door face are arranged asymmetrically and 2 coils are removed on the face of the MSR opposite the door because a projector porthole blocks their positioning). We used the MC elements to generate known fields by which to calibrate the OPMs. However, to do this we need to know what field each coil element produces – this is termed the MC forward model.

*2.3.1) Determining the MC forward model*

One could in principle use the positions of the wire paths of each coil to generate a forward model to ensure accurate knowledge of the fields generated by each MC element. However, interactions between the wire paths and the MuMetal walls of the MSR (Hammond, 1960; Roshen, 1990), complexities in monitoring many coil currents, and potential inaccuracies involved in mounting the coils means we chose instead to use a data-driven approach, similar to that originally described by Iivanainen et al. (2022). Briefly, Iivanainen et al. (2022) measured the fields produced from a system of 18 coils placed inside a cylindrical magnetic shield (MSC) over a volume containing an OPM-MEG helmet. By fitting the fields to a vector spherical harmonic model, they obtained a forward model describing the field generated by each coil, within a volume of interest, per unit of applied current. This approach has also been used to measure remnant fields inside MSRs (Holmes et al., 2022; Mellor et al., 2021; Rea et al., 2021) and to model both neuronal and interference fields measured by MEG systems (Taulu and Kajola, 2005; Tierney et al., 2022, 2021).

Here, to derive the data for the harmonic model, we used a triaxial Bartington Mag-13MSL100 fluxgate magnetometer (Bartington Instruments, Witney, UK) integrated with an optical tracking system (Natural Point Inc., Corvallis, OR, USA) which uses infrared light to determine the position of retroreflective markers placed onto the fluxgate. (The use of multiple markers enables measurement of the sensor orientation) (see Rea et al. (2021) for details). We used a wooden board and marked a 40 x 40 cm$^2$ plane in 10-cm intervals which could be raised or lowered using 10-cm spacers to cover a 40 x 40 x 40 cm$^3$ volume. For each of the 125 grid positions we placed the magnetometer in position, used the optical tracking system to get its position and orientation and then drove each of the 94 coils at a different frequency. We chose frequencies between 2 and 11.5 Hz with a separation of 0.1 Hz, recording for 20 s for sufficient frequency resolution (0.05 Hz) to resolve all peaks. We used a NI-cDAQ-9179 (National Instruments) to control both the series of NI-9264 16-bit digital to analogue converters which generated the voltage signals applied to the coil drivers and the NI-9205 analogue to digital converters which collected the data. We used the MATLAB NI-DAQmx package function 'readwrite' to simultaneously generate and record signals with a sampling frequency of 10 kHz. Following data collection, we moved the fluxgate to the next position until the whole volume had been sampled. Once data had been collected, we extracted the direction and magnitude of the field generated at each grid point by each coil. We then fitted these data to a 4$^{th}$-order regular spherical harmonic model. The RMS error between the data and the model was <1% for all coils.

*2.3.2) OPM Calibration procedure*

To calibrate the OPMs, we used the same signal used to derive the MC forward model, with a maximum field generated of around 1,000 pT. We took the FFT of the resulting OPM data, to quantify the field detected by the OPMs from each coil (based on the known coil frequencies) and minimised the difference between the measured data ($v_n$ at channel n) and the forward model to determine sensor locations orientations and gains. Mathematically,

$$\min_{r,\hat{o},g} \left\| \sum_{n=1}^{N} v_n - g \boldsymbol{B_n}(\boldsymbol{r}) \cdot \hat{\boldsymbol{o}} \right\| \qquad [2]$$

for each channel. Here $\boldsymbol{B_n}(\boldsymbol{r})$ denotes the field from each matrix coil from the MC forward model. We reduced the number of coils used from 94 to 12 (2 randomly selected from each face) and set a (empirically derived) threshold level for optimisation and randomised the selected coils in the system until this threshold was met, or 100 iterations were completed (typically this indicated an inactive sensor or a channel with high noise). This was done as using all coils together made the system perform poorly, possibly as the initial search space is difficult to navigate with many coils. We again assumed in our minimisation that each of the channels in an individual OPM sensor shared the same location and optimised the three channels of each sensor simultaneously. The initial guess for OPM position was the centre of the grid used to map the coils, with the channels again oriented along the approximate front-back alignment direction of the helmet with 2.7 V/nT gain. We note that Iivanainen et al. first used their coils to generate three orthogonal uniform fields and the complete set of five field gradients to obtain an initial guess of the sensor orientation and then position before refining with the harmonic model as above. We found that this initial step was not needed after reducing the number of coils used in the minimisation. This may be due to the larger distance between our MC and the helmet (1.4 m) compared to their experimental setup.

**2.4) Validation experiments**

To validate our two calibration methods (HALO and MC), we undertook a series of experiments. For each experiment, the OPMs were located in the 3D printed helmet (Figure 1C). All sensors were used 'as is' from the production line (i.e., with no in factory calibration). This means that gain and orientation sensitivity are subject to errors from manufacturing tolerances and operational settings as well as local conditions. During the sensor start-up procedure, we used the manufacturer's software to perform a "field zeroing" process, by which the on-board sensor coils were used to offset any remnant DC background field, ensuring a close to zero field within the cell for the experiment. The manufacturer's software also carries out an additional step in which each of the three on-sensor coils is used to produce a field of known amplitude. Measurements of the amplitude and phase of the variation of transmitted beam intensity at the modulation frequency for the three different applied fields are then used to establish signal weightings that yield values of the field components oriented along the three orthogonal directions defined the three sensor coil fields (which are approximately equivalent to the orientations of the sides of the sensor casing). These measurements are used to set the assumed sensor gain to 2.7 V/nT for all three axes. Following this, we performed our calibration procedures, as described above.

A single experiment comprised calibration with both the HALO and MC. This was repeated 11 times. For the first 5 repeats, the MSR was opened, and the HALO moved to a different position relative to the helmet (specifically it was positioned on the top, left, right, back and front of the helmet, as shown in Figure 3 and remained there throughout the experiment). For the remaining 6 repeats nothing was changed between experiments (and the HALO was sited on the top of the helmet). The OPMs were always restarted between successive experiments. The order in which the HALO and MC calibration occurred was switched for each experiment.

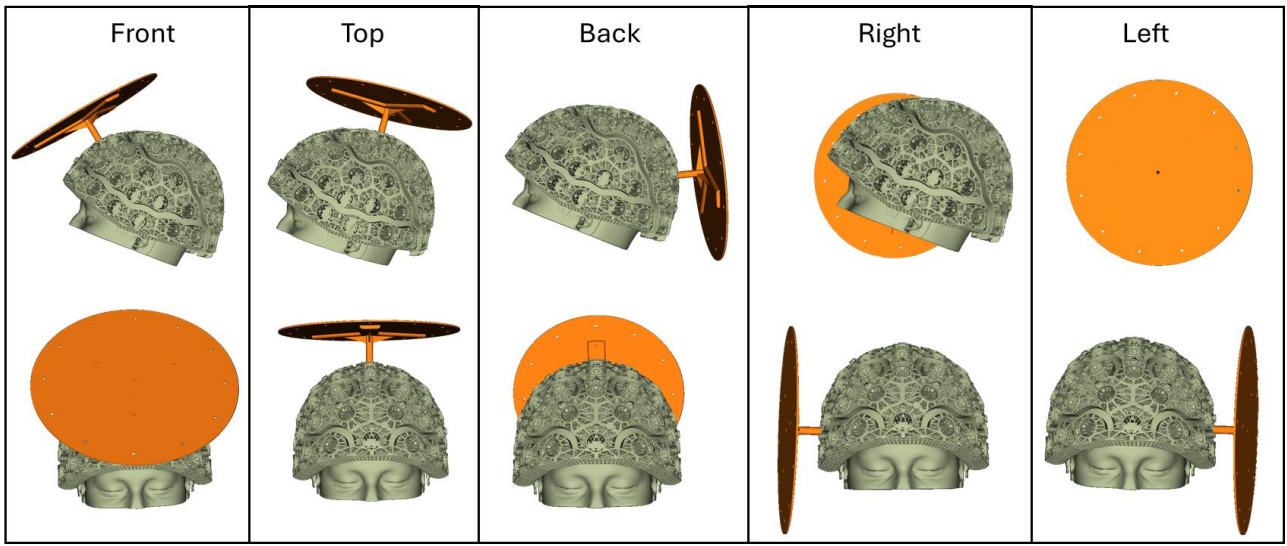

*Figure 3: HALO positioning:* All 5 locations the HALO was placed within the helmet.

Because the HALO was inserted into one of the sensor slots in the helmet. We had to remove one sensor from the helmet, and hence the calibration process. We also removed any sensors with excessive noise levels. This left us with seven datasets with 63/64 sensors, two datasets with 62/64 sensors and two datasets with 61/64 sensors. Calibration was also made complicated since, by moving the location of the HALO, we were not able to repeat calibration for every possible sensor slot in the helmet, for every experiment performed. For this reason, some slots had more repeat experiments (ranging from 11 to 4 repeats).

Following data collection and fitting (using the procedures described above) for each of the 11 runs, we had two independent system calibrations (one from the MC array and one from the HALO), both comprising sensor locations, orientations and values of the gain for each channel. We quantitatively compared these data in three ways:

1) **Location**: The only 'ground truth' available was the locations of the cell for each OPM, which are known from a combination of the helmet CAD file and the design of the OPM. The accuracy of these values is limited by the manufacturing tolerance of the helmet and OPM, the precise location at which the laser passes through the cell (which we assumed to be the centre), and the extent to which all sensors are correctly inserted into the helmet. Nevertheless, we expect these CAD-derived locations to be accurate to within ~2 mm. Comparison of these values to the calibration-derived locations is somewhat complicated since the CAD locations are in the coordinate space of the helmet, and the calibration locations are in the coordinate space of either the HALO or the MC. To quantitatively compare the two, we first computed the Euclidean distances from every sensor in the helmet CAD model to every other sensor. We repeated this calculation for the locations derived using both calibration methods and computed a matrix of residuals by taking the absolute value of the difference between the CAD-derived distances and the calibrated distances. We calculated the mean and standard deviation of these values for each sensor across experiments to form an estimate of the accuracy and repeatability of the location estimate for each sensor. In addition, we also used an iterative closest point (ICP) algorithm to perform a rigid body transformation of the calibration-derived sensor locations to the CAD locations. The ICP algorithm minimised the mean Euclidean distance between CAD and calibration-derived locations. We then took the final mean Euclidean distance as an *absolute* value of the difference between the calibration and CAD-derived locations.

2) **Orientations**: The assumed sensitive orientation of OPM measurements relates to the modulation field directions, which in turn depend on the orientation of the on-board sensor coils. These orientations should point along the axes of the casing of the OPM (see Figure 1A). We again used the ICP algorithm to rotate/translate the calibration-derived locations to the CAD locations. We then compared the three orthogonal axes of the sensor casing (from the CAD) to the calibration-derived orientations for each sensor axis. Specifically, for all three axes and for all OPMs we measured the angle between the assumed axis and calibration-derived axis. We then compared both the stability of each method, and the agreement between methods, by calculating and plotting the mean and standard deviation across all available runs from HALO-derived angles against MC-derived angles. We reasoned that if both approaches were stable, the error-bars would be small for all axes (i.e., the angle difference similar for each run). We also reasoned that a linear relationship would suggest that the calibration methods were equivalent (i.e. we find the same sensor orientation with the HALO and the MC calibration).

3) **Gains**: The gain of each OPM channel is assumed to be set using the sensor start-up procedure. We therefore expected our recovered gains to be centred around 2.7 V/nT. We produced equivalent plots to those generated for the orientations to compare the stability of and agreement between methods.

**2.5) Human experiments: data acquisition**

We undertook a series of human experiments with the aim to determine whether calibration of the OPMs leads to an improvement in OPM-MEG performance when assessing data in source space. Seven healthy participants (2 identified as female, 5 identified as male, mean age 32 ± 7 years, all right-handed) took part in the study, which was approved by the University of Nottingham Faculty of Medicine and Health Sciences Research Ethics Committee (approval number H16122016). All participants provided written informed consent.

All subjects underwent a visuo-motor experiment designed to elicit oscillatory responses in both the beta and gamma bands, as well as evoked responses. The experimental paradigm incorporated two types of trials:

1) **Circles trials:** A visual stimulus (a central, inwardly moving, maximum-contrast circular grating) was presented. The grating subtended a total visual angle of 14° and was displayed for 1.5 s. This was followed by a baseline period lasting 2 s, during which a central fixation cross was displayed. Whilst the circle was on the screen the subject was asked to make repeated abductions of their right index finger. There were 60 circles trials per experiment. This visual stimulus is known to induce gamma oscillations in visual cortex (Hoogenboom et al., 2006) whilst the finger movement is known to modulate beta oscillations in motor cortex (Pfurtscheller and Lopes da Silva, 1999).

2) **Faces trials:** In these trials a photograph of a face was displayed on screen for a duration of 0.3 s, followed by a (jittered) rest period of duration $1 \pm 0.1$ s (during which a central fixation cross was shown). A total of 180 faces trials was used. This task generates evoked responses in primary visual and fusiform areas (Bentin et al., 1996; Halgren et al., 2000; Hill et al., 2022; Taylor et al., 2001).

In all cases, visual stimuli were presented via projection through a waveguide in the MSR onto a back projection screen, positioned ~100 cm in front of the subject (using a ViewSonic PX748-4K data projector).

In an experimental session, the subject was first seated on a patient support at the centre of the MSR. The OPM helmet was placed on the subject's head, and the HALO was positioned on the top of the helmet (secured via an empty OPM slot – Figure 3 (top)). Following this, the MSR door was closed, and the inner walls of the room degaussed (this process took 1 minute). The manufacturer field-zeroing and parameter setting was performed and then calibration of the OPMs was performed using both the HALO and the MC; this process took a total of 2 minutes. The experimental paradigm was then run, and OPM-MEG data collected at a sample rate of 375 Hz. The experimental paradigm lasted ~550 s and the total experimental time (including degaussing, calibration using both techniques, and the experiment itself) was ~13 minutes.

Immediately following MEG data acquisition, we used an optical scanning technique to determine how the helmet was positioned on the subject's head. Specifically, a 3D digitisation of the participant's head (with the helmet in place) was acquired using a 3D structured light scan (Einscan H, SHINING 3D, Hangzhou, China). The 3D surface of the subject's face was extracted from this scan and matched to the equivalent surface taken from a T1-weighted anatomical magnetic resonance image (MRI). This enabled a co-registration of the helmet relative to brain anatomy (Hill et al., 2020; Zetter et al., 2018).

**2.6) Data analysis**

OPM-MEG data for every channel were initially inspected by computing the power spectral densities, and any channels with a high noise floor (taken as >30 fT/√Hz in the 60 – 80 Hz band) or very low signal levels (<7 fT/√Hz) were removed. A trial-by-trial analysis was also carried out, whereby trials with variance greater than 3 standard deviations from the mean trial variance were automatically removed. Data were also inspected visually, and any obvious noisy channels or trials removed. A 1 – 150 Hz band pass filter was applied along with notch filters at the powerline frequency (50 Hz) and two of its harmonics.

We used a beamformer spatial filter (Robinson and Vrba, 1998) to process data in source space. This requires accurately calibrated data, and we used 3 separate approaches:

1) **Assumed calibration**: Sensor locations were derived from the helmet CAD; the location of the helmet relative to the brain was derived from the structured light scan and MRI. The combination of these data allowed co-registration of sensor locations relative to the brain. The orientations were assumed to be orthogonal, and parallel to the sides of the sensor casing. The sensor gain was assumed to be 2.7 V/nT.

2) **HALO calibration**: We used the ICP algorithm to co-register the HALO-derived locations to the helmet CAD, and this coupled with the structured light scan and MRI allowed sensor locations relative to the brain to be derived. For every channel, the orientation and gain were taken directly from the HALO fitting algorithm (with orientations rotated to the same space as the MRI scan). Recorded voltage data for each channel were divided by the calculated gain value for each channel to yield field data.

3) **MC calibration**: This was identical to the HALO calibration but using MC-derived coefficients (locations, orientation, and gains).

Notice that for all three calibrations, the processes of co-registration was identical. Having calibrated the system in these three ways, a forward model was constructed using a single shell volume conductor model (Nolte, 2003). Following this, we undertook three separate analyses:

**Gamma modulation:** Data collected during the circles trials were segmented to 0 s to 3.5 s windows (relative to the onset of the circle) and filtered to the 35 Hz to 55 Hz band. A covariance matrix and beamformer weights were constructed using these data. To make an image showing the spatial signature of stimulus induced gamma change, we contrasted 35-55 Hz oscillatory power in the 0.5 s to 1.5 s (active) window to power in the 2.5 s to 3.5 s (control) window, deriving a pseudo-T-statistic (Vrba and Robinson, 2001) for all voxels on a regular 4-mm grid covering the brain. Source orientation for each voxel was taken as that with the largest signal power. Having derived the location of largest gamma change, we then constructed a 'virtual electrode' showing the timecourse of electrophysiological change at this peak location. This was done in two ways; first, we generated a gamma amplitude envelope timecourse by taking band-limited data for each trial, applying a Hilbert transform to get the analytic signal, and then taking the absolute value of the analytic signal and averaging over trials (this is termed the Hilbert envelope). Second, we recalculated the beamformer weights using broadband (1 – 150 Hz) data, and then constructed a time frequency spectrogram (TFS) by sequentially filtering data into overlapping frequency bands, deriving the Hilbert envelope for each band and concatenating these data in frequency. In both cases, an estimate of baseline amplitude was derived (independently for each frequency band) in the control window (3 s to 3.5 s) and this was subtracted; the data were also normalised by these same baseline values to give an estimate of change relative to baseline 'noise'. Finally, gamma band signal-to-noise ratio (SNR) was measured as the change in mean gamma amplitude (between the active and control window) divided by the standard deviation of the envelope in the control window.

**Beta modulation**: Stimulus induced change in beta power (due to finger movement) was mapped in a similar way to the gamma change. Data from Circles trials were segmented into 6-s windows (starting at the onset of the circle stimulus), filtered into the 13 Hz to 30 Hz band, and covariance and beamformer weights calculated. A pseudo-T-statistical image was generated, contrasting active (1 s to 2 s) and control (2.5 s to 3.5 s) windows. A Hilbert envelope of beta amplitude and a TFS were derived for the location of maximum beta modulation as described above (but with a baseline window of 5 s to 6 s), and SNR was calculated as the difference in mean beta amplitude in the active and control windows divided by the standard deviation in the active window.

**Evoked responses:** To look at evoked responses to faces, data were segmented to -0.2 s to 1.3 s time windows (relative to the presentation of a face) and frequency filtered to the 2 Hz to 40 Hz band. Covariance and weights were constructed using data from all faces trials. To compute the evoked response, we first used a beamformer to reconstruct a virtual electrode at an anatomically defined point in the primary visual cortex (selected according to the automated anatomical labelling atlas). Beamformed timecourses were averaged across trials, giving the evoked response. For the peak in the evoked response at 150 ms, we generated a pseudo-Z-statistical image (which contrasts beamformer projected source amplitude (at a single point in time) to the noise amplitude (Vrba and Robinson, 2001)). The instantaneous beamformer projected amplitude was normalised by an estimate of the projected noise (taken as $(\boldsymbol{w}^T \boldsymbol{C}_N \boldsymbol{w})^{1/2}$ where $\boldsymbol{w}$ are the beamformer weights and $\boldsymbol{C}_N$ is the 2 – 40 Hz covariance estimated in a control window (1.1 s to 1.3 s)). Evoked response SNR was estimated as the maximum amplitude of the evoked response divided by the same projected noise.

Each of the above three analyses was undertaken (independently) for each of the three calibration types (assumed (CAD), HALO and MC). We tested statistically (using Wilcoxon signed rank test) whether SNR was increased when using either the HALO or MC compared to the assumed calibrations. We also explored the source localisation repeatability by comparing the peak location for each participant in each method. To do this, we co-registered each subject to the MNI-152 average brain using FLIRT (FSL) (Jenkinson et al., 2002; Jenkinson and Smith, 2001) and computed the average and standard deviation of the peak location across participants for each case. We then calculated an "error ellipsoid" (one for each calibration method) that was centred on the average location and with each axis of the ellipsoid representing the standard deviation of peak locations in X, Y, and Z across participants. We reasoned that if all three methods localised with similar accuracy, the ellipsoids would overlap with similar volumes (suggesting that errors are dominated by intersubject variation rather than calibration method).

## 3) RESULTS

### 3.1) Validation experiments

Figure 4A shows the basic output of the calibration procedures. In both the left and right panels, the grey data-points and arrows show the sensor locations and sensitive orientations, according to CAD, respectively. In the left-hand plot, the coloured data-points and arrows show the HALO-derived locations and orientations following the ICP process. In the right-hand plot, the coloured data points and arrows show the MC-derived locations and orientations following the ICP process. In both cases, the colour of the arrow represents the derived gain multiplier for that channel (i.e., relative to the assumed value of 2.7 V/nT); blue shows gain scaling values <1 and red shows gain scaling values >1. Note that grey arrows are used for the CAD data as we always assume a gain of 2.7 V/nT. Data from a single representative experimental run are shown. For this specific run, the mean (across sensors) *absolute* Euclidean distance from the CAD-derived to the HALO-derived sensor location was 2.9 mm; the equivalent distance from the CAD to the MC-derived location was 3.4 mm. Across all experimental runs, the average a*bsolute* Euclidean distances were 3.0 ± 0.1 for the HALO and 3.7 ± 0.3 for the MC (mean ± standard deviation values shown). These values show good agreement between calibration and the CAD (the latter being the closest we have to a ground truth).

A limitation with these absolute discrepancies is that they do not allow investigation of which sensor locations are best matched to the CAD-derived locations. This is because our ICP algorithm minimises the average distance between CAD- and calibration-derived locations, and this could mask individual sensors that offer only a poor agreement to CAD. For this reason, we also derived sensor-to-sensor distances across the helmet (which are independent of the ICP algorithm). Figures 4B, C and D show matrices representing the residuals between the CAD-derived and the calibration-derived sensor-to-sensor distances. Figure 4B shows the comparison between the HALO and CAD. Figure 4C shows the comparison between the MC and CAD. Figure 4D shows the comparison between the HALO and the MC. Note first that the residual matrices in Figure 4B and Figure 4C appear visually similar (the Pearson correlation coefficient between these matrices is 0.77). This means that sensors in which HALO-derived locations differ most from CAD, also tend to be the sensors where the MC-derived locations differ most from CAD. In general, the residuals between the HALO and MC sensor-to-sensor distances were lower, suggesting that the HALO and MC are in better agreement with each other than they are with CAD. This suggests the dominant inaccuracies relate to the CAD itself (i.e. helmet manufacture and sensor placement).

Formalising the above results, the average residual sensor-to-sensor distance (across all experimental repeats) was 2.08 ± 0.07 mm for HALO vs. CAD; 2.62 ± 0.19 mm for MC vs. CAD, and 1.97 ± 0.16 for HALO vs. MC. (All cases show mean ± standard deviation.) Most importantly, these values, and the matrices from which they are derived, show that both of our calibration techniques are able to localise sensors to within a few millimetres of their CAD-derived locations, and that these residual 'errors' most likely derive from the CAD.

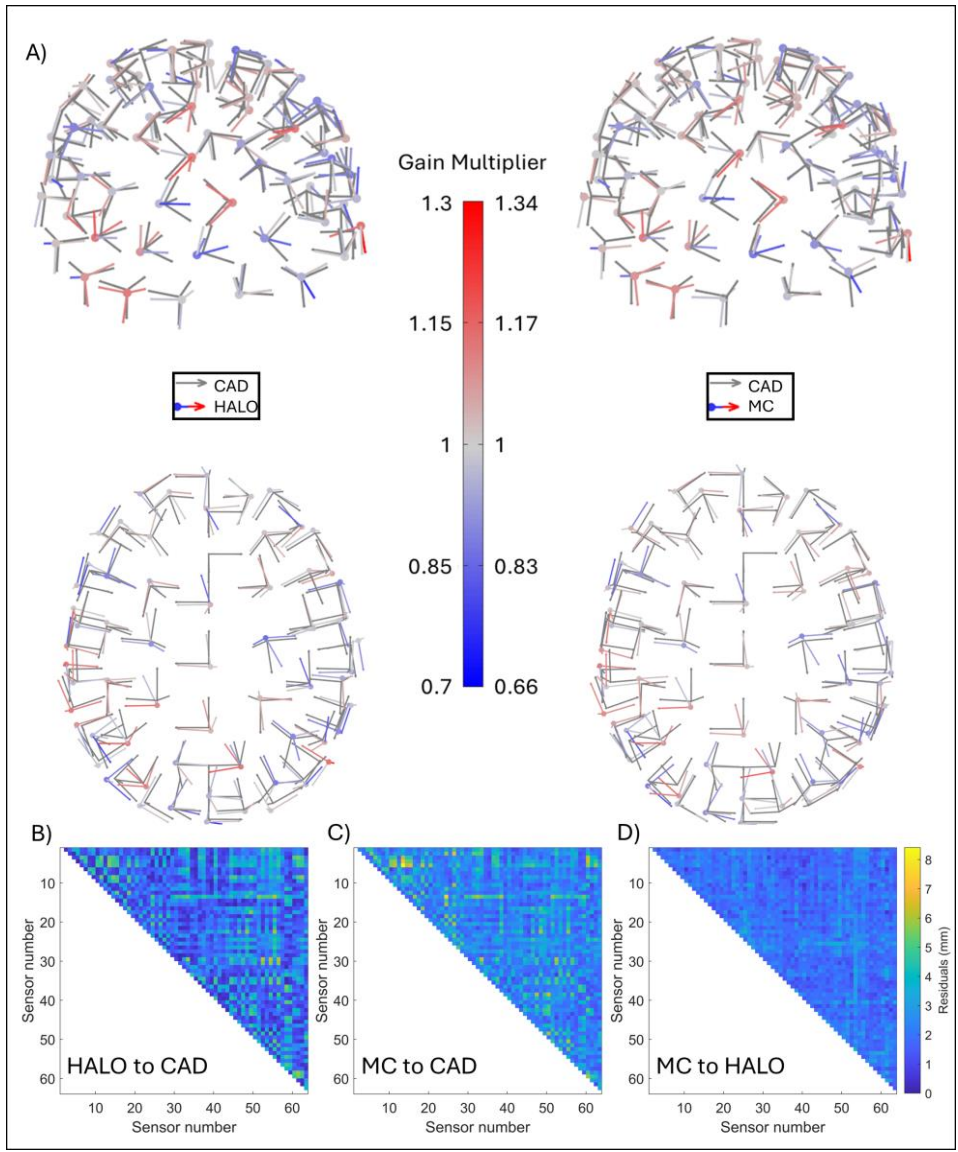

*Figure 4: Quantifying the accuracy of calibration-derived sensor locations. A) CAD- (grey) and calibration- (coloured) derived sensor locations, orientations and gains. Left panel shows HALO, right panel shows MC. In both cases, the data points show the sensor location, the arrows show three sensitive orientations, and the colour shows the gain multiplier. (in the case of the CAD-derived values the gain multiplier is assumed to be 1). B) Matrix showing the mean residual sensor-to-sensor distance between CAD- and HALO-derived locations. C) Matrix showing the mean residual sensor-to-sensor distance between CAD- and MC-derived locations. D) Matrix showing the mean residual sensor-to-sensor distance between MC- and HALO-derived locations.*

Figure 5 shows how our calibration-derived estimates of the sensitive orientations of the OPM differ from the CAD models (these differences are measured as angles, so an angle of 10° means the orientation of the sensitive axis in the OPM differs by 10° from the orientation of its outer casing). Figures 5a-c plot the MC-derived angular discrepancy against the HALO-derived angular discrepancy. Panels A, B, and C show the x, y, and z axes of each OPM, respectively. In all three cases, the data points show the mean angular discrepancy, for a single OPM, calculated across experimental runs. The error bar shows the standard deviation across runs, meaning a small error bar denotes good consistency in the calibration-derived orientation across experiments. In all cases, the best fit line is shown in red and the line of equality in black. The fact that these lines fall very close to each other indicates good agreement between methods (i.e., the sensitive orientation derived using the HALO is similar to that derived using the MC).

For each axis the mean, standard deviation, minimum, and maximum angular discrepancies across all runs and channels were: x-axis: 17.1 ± 8.6° (max/min = 37.1/2.0°); y-axis: 5.4 ± 2.3° (max/min = 10.7/0.3°) and z-axis: 10.3 ± 5.3° (max/min = 23.1/0.8°) using the HALO and x-axis: 17.2 ± 8.5° (max/min = 36.0/2.4°), y-axis: 5.4 ± 2.4° (max/min = 11.3/0.2°) and z-axis: 10.5 ± 5.4° (max/min = 23.7/1.3°) using the MC. The x-axis has the largest error and widest range of values and Y-channel the smallest error and narrowest range of values. The wide discrepancy across both sensors and channels with the CAD model is maintained across methods, highlighting the importance of calibration.

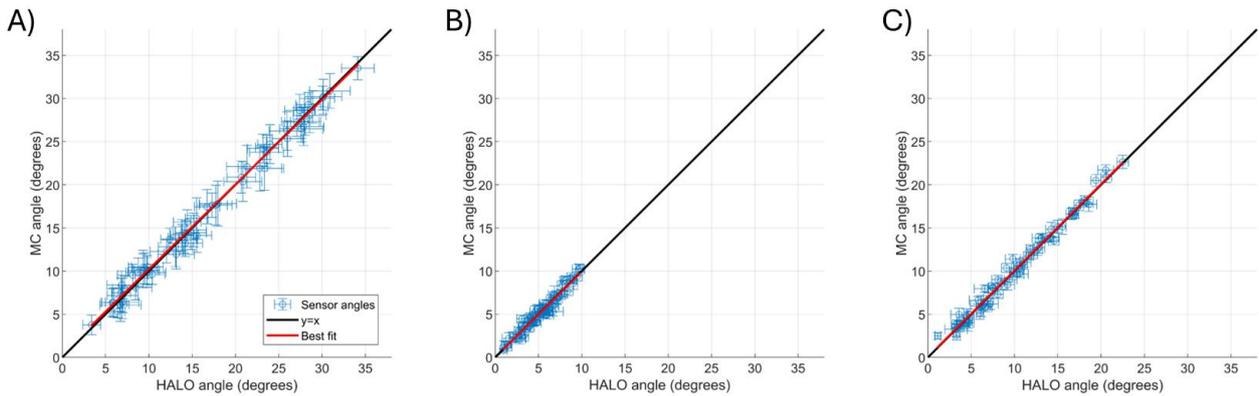

*Figure 5: Estimating the accuracy of calibrated sensor orientations. All panels show the MC-derived angular discrepancy (with respect to CAD) against the HALO-derived angular discrepancy (also with respect to CAD). i.e. an angle of 10° would mean that the orientation of the sensitive axis in the OPM differs by 10° from the orientation of the outer casing of the sensor. Panels A, B and C show the x, y, and z axes of each OPM, respectively. Red lines show a best fit to the data. Black lines show the line of equality. All data points show the mean value across experimental runs and the error bars represent standard deviation. A close linear relationship suggests that the two calibration methods result in very similar sensitive orientations.*

Figure 6 shows equivalent plots to Figure 5 but for the derived gain values. All values are centred around the expected 2.7 V/nT gain and again the best-fit line closely follows the line of equality, showing that the two calibration techniques generate similar gain values, for each sensor. For each axis the mean, standard deviation, and minimum and maximum values across all runs and channels were x-axis: 2.61 ± 0.24 V/nT (max/min = 3.15/2.01 V/nT), y-axis: 2.78 ± 0.19 V/nT (max/min = 3.75/2.39 V/nT) and z-axis: 2.67 ± 0.17 V/nT (max/min = 3.01/2.23 V/nT) for the HALO and x-axis: 2.63 ± 0.25 V/nT (max/min = 3.16/2.13 V/nT), y-axis: 2.80 ± 0.19 V/nT (max/min = 3.71/2.42 V/nT) and z-axis: 2.69 ± 0.16 V/nT (max/min = 3.07/2.29 V/nT) using the MC. Again, a discrepancy between the recovered values and the expected value highlights the importance of calibration and the accuracy of both methods.

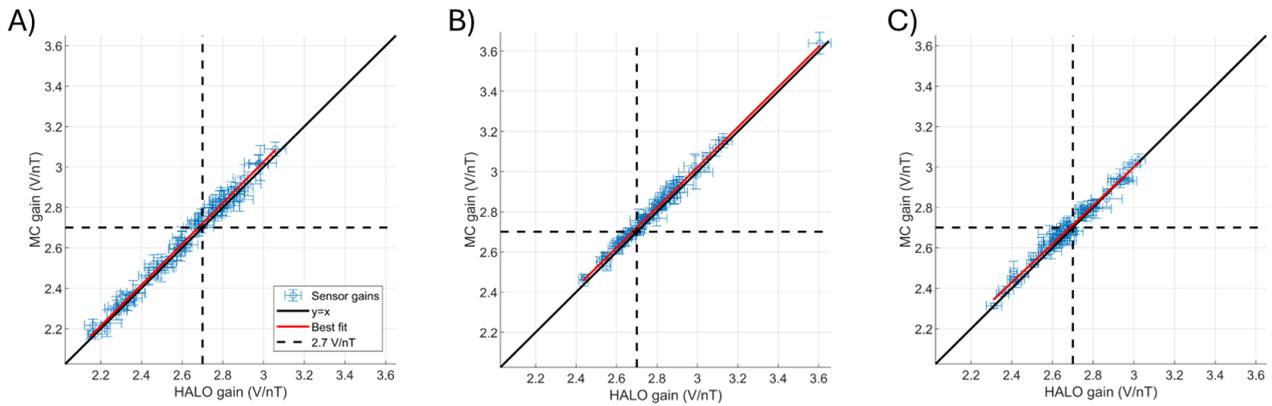

*Figure 6: The accuracy of the calibrated sensor gains compared with the expected value (of 2.7 V/nT). Each panel plots the HALO-recovered gain against the MC-recovered gain. The values are shown as the mean across repeats, with the standard deviation shown by the error bars. Panels (a), (b) and (c) show angles for the sensor X, Y and Z channels respectively. The dashed crosshairs show the expected gain value of 2.7 V/nT.*

**3.2) Human experiments**

Figure 7 shows the beta band modulation induced by finger abduction. Figure 7A shows the group-averaged pseudo-T-statistical image, derived independently using the CAD- (top), HALO- (middle), and MC- (bottom) derived calibrations. The colour overlay, which shows the spatial signature of task induced beta power loss during movement, is thresholded to 80% of its minimum value. In all cases, the largest beta modulation is observed in the sensorimotor areas as would be expected. However, the pseudo-T value itself (a proxy metric for SNR) is 'largest' (i.e. most negative) for the HALO-derived calibration (-0.41 ± 0.03 – mean ± standard deviation over subjects); the MC-derived calibration resulted in a marginally lower pseudo-T value (-0.37 ± 0.03), and the CAD-derived calibration shows the smallest (least negative) value (-0.29 ± 0.04). This suggests a better reconstruction of the source for both the HALO and MC calibration, compared to the CAD model, with an average increase of 1.42 ± 0.09 and 1.31 ± 0.09 respectively.

Figure 7A also shows the group-averaged TFS for a source at the location of the peak pseudo-T-statistic, for the three calibrations. The three plots are similar, with the expected movement-related beta decrease followed by an increase above baseline (the beta "rebound") immediately following movement cessation. The TFSs reconstructed using either the HALO or MC calibrations show larger relative change from baseline compared to the TFS produced using CAD calibration. This difference is quantified in Figures 7B/C. Figure 7B shows the beta-band envelope averaged over participants, with the standard deviation indicated by the shaded area. Figure 7C shows the reconstructed source-space SNR for each subject. While both calibration methods show significant (p = 0.0156 for both methods, Wilcoxon sign-rank test) increases in SNR compared to the CAD-derived calibration, the HALO-derived calibration provides a greater increase. On average, the SNR increased by a factor of 2.85 ± 0.95 for the HALO and 1.77 ± 0.23 for the MC-derived calibration.

Figure 7D shows the location of maximum beta change for each subject, after transformation to the MNI-152 average brain. The average reconstructed location and its standard deviation in each axis are represented by the shaded ellipsoids. For all participants, the activity is localised to the precentral gyrus and the error ellipsoids overlap. However, for Subject 2 in the CAD-derived data, the peak location was found in the right precentral gyrus (i.e., an ipsilateral response) – the HALO- and MC-derived calibrations did not show this. The ellipsoids have volumes of 6417 mm$^3$, 829 mm$^3$, and 889 mm$^3$ for the CAD, HALO, and MC calibrations respectively, suggesting that the CAD-derived peak locations have a greater spatial spread while the HALO and MC are similar. These peak localisations are summarised in Table 1.

Figure 8 shows gamma band (40 – 60 Hz) oscillations induced by visual stimulation using the circular grating. (The Figure layout is equivalent to Figure 7.) The pseudo-T statistics at the locations of maximum gamma change were 0.48 ± 0.30 for the CAD calibration, 0.68 ± 0.47 for the HALO, and 0.62 ± 0.42 for the MC. Again, this suggests a better source reconstruction with the HALO and MC, with an average increase of 1.32 ± 0.17 and 1.22 ± 0.15 respectively. The TFS again show larger relative change when calibration is carried out using either the HALO or the MC compared to CAD. On average, the SNR increased by a factor of 1.32 ± 0.35 for the HALO and 1.25 ± 0.41 for the MC-derived calibrations. However, whilst the improvement in SNR was significant for the HALO at the group level (p = 0.047 - Wilcoxon sign rank test) it was not significant for the MC (p = 0.38 – Wilcoxon sign rank test). Data in Figure 8C show that SNR improvements were less pronounced for gamma activity than they were for beta-band activity. Indeed, for some subjects SNR was marginally reduced – this will be addressed further in the Discussion. For all participants, using all three calibrations, the maximum gamma modulation activity was localised to primary visual cortex, as shown in Figure 8D. The ellipsoids have similar volumes of 822 mm$^3$, 703 mm$^3$, and 776 mm$^3$ for the CAD, HALO, and MC calibrations respectively.

Figure 9 shows the results of recording evoked responses to faces. Figure 9A shows the subject averaged (normalised) pseudo-Z-statistical map at a time t = 150 ms, thresholded to 80% of its maximum value. All three cases show an average localisation in the occipital lobe. The activity derived using the CAD data localises entirely to the calcarine; for the MC-derived calibration, activity localises to the cuneus; and the HALO-derived activity bridges the two regions. Figure 9B shows the reconstructed timecourse for a point at the centre of mass of the calcarine (as determined by the AAL atlas for each participant). For the three calibrations, timecourses show similar temporal morphology, with the HALO- and MC-derived positions showing larger peaks at 120 ms and 150 ms compared to the CAD data. Figure 9C shows the SNR for all participants, which increased significantly for both the HALO and MC compared to CAD (p = 0.0156 for both methods – Wilcoxon sign rank test). On average, the SNR increased by a factor of 1.31 ± 0.19 for the HALO and 1.21 ± 0.10 for the MC-derived calibrations. Figure 9D shows that, for all participants, the peak at 150 ms was localised to the occipital lobe with the mean location (i.e. the average across participants) remaining similar in each case. However, there was a greater standard deviation of peak location in the CAD-derived case. The ellipsoids have volumes of 13,363 mm$^3$, 3,808 mm$^3$, and 4,900 mm$^3$ for CAD, HALO, and MC calibrations respectively. This again implies a larger scatter of locations for the CAD-derived peak.

| Method | Mean Position (mm) | | | Standard Deviation (mm) | | | Volume (mm$^3$) | | |
|---|---|---|---|---|---|---|---|---|---|
| | Beta | Gamma | Evoked | Beta | Gamma | Evoked | Beta | Gamma | Evoked |
| CAD | [-26, -13, 64] | [-2, -92, 5] | [9, -91, 25] | [31, 8, 5] | [7, 4, 5] | [19, 10, 15] | 6417 | 822 | 13363 |
| HALO | [-40, -16, 60] | [-4, -93, 6] | [0, -91, 21] | [6, 7, 4] | [5, 5, 5] | [10, 5, 14] | 829 | 703 | 3808 |
| MC | [-39, -17, 63] | [-4, -92, 5] | [0, -91, 23] | [7, 7, 3] | [4, 5, 7] | [10, 7, 16] | 889 | 776 | 4900 |

*Table 1: **Localisation summary** – The mean and standard deviation for peak localisation across participants for each of the three tasks using each of the three calibration methods. The final column shows the volumes of the error ellipsoids plotted in Figures 7D, 8D, and 9D.*

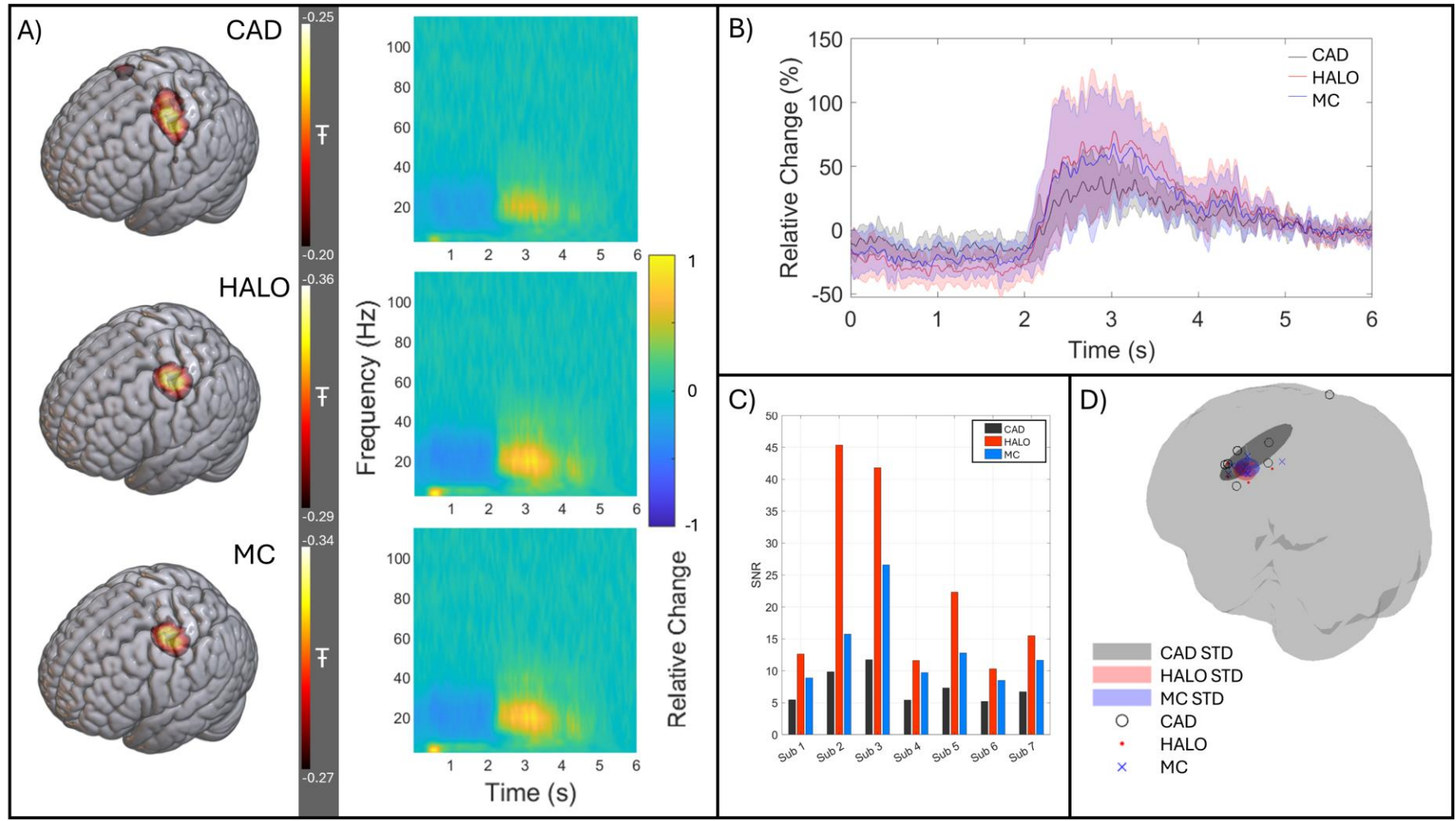

*Figure 7: Beta band effects* – A) Average pseudo-T-statistical image (thresholded at 80% of their minimum value) and a TFS showing the evolution of neural oscillatory activity at the location of peak beta modulation; data averaged across all participants for the CAD- (Top), HALO- (Middle), and MC- (Bottom) derived calibrations. B) The source reconstructed beta-band envelope for each method (CAD, HALO and MC in black, red and blue respectively) averaged across subjects with the standard deviation shown as a shaded area. This was again calculated at the location of largest beta modulation for each subject. C) Beta-band SNR calculated for each subject for CAD-, HALO-, and MC-derived calibration in black, red and blue respectively. D) The location of largest beta change for each participant, using all three calibration methods. CAD peaks are shown as black circles, HALO as red dots, and MC as blue crosses. In addition, the mean position and standard deviation in each axis are shown by shaded ellipsoids.

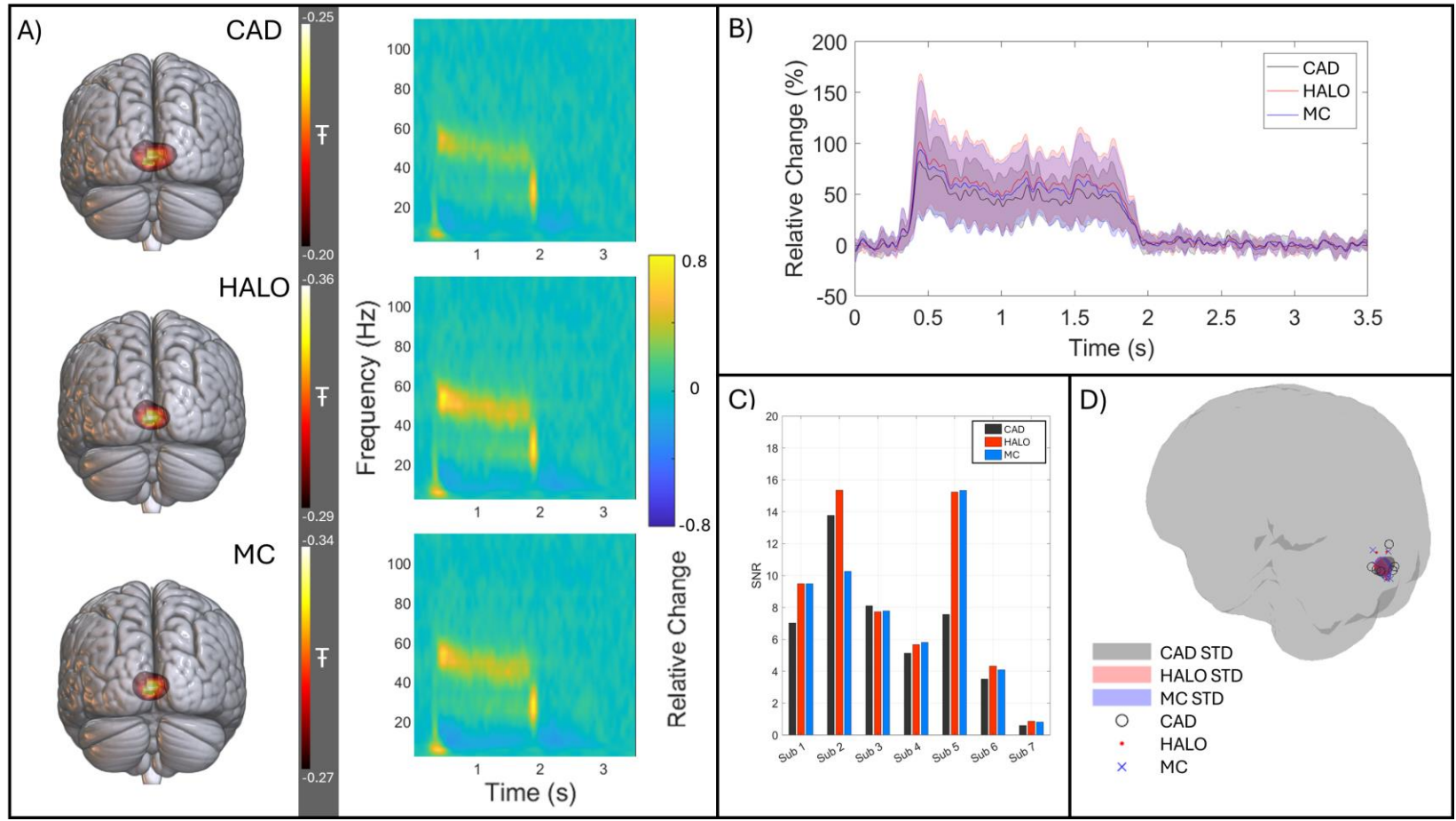

*Figure 8: Gamma-band effects* – A) Average pseudo-T-statistical images (thresholded at 80% of their maximum value) and a TFS showing the evolution of neural oscillatory activity at the location of peak gamma modulation; data averaged across participants for the CAD- (Top), HALO- (Middle), and MC- (Bottom) derived calibrations. B) The source reconstructed gamma-band envelope for each calibration method (CAD, HALO and MC in black, red and blue respectively) averaged across subjects with the standard deviation shown as a shaded area. This was again calculated at the location of largest gamma modulation for each subject. C) Gamma-band SNR calculated for each subject for CAD-, HALO-, and MC-derived calibration in black, red and blue respectively. D) The location of largest gamma change for each participant, using all three calibration methods. CAD peaks are shown as black circles, HALO as red dots, and MC as blue crosses. In addition, the mean position and standard deviation in each axis are shown by shaded ellipsoids.

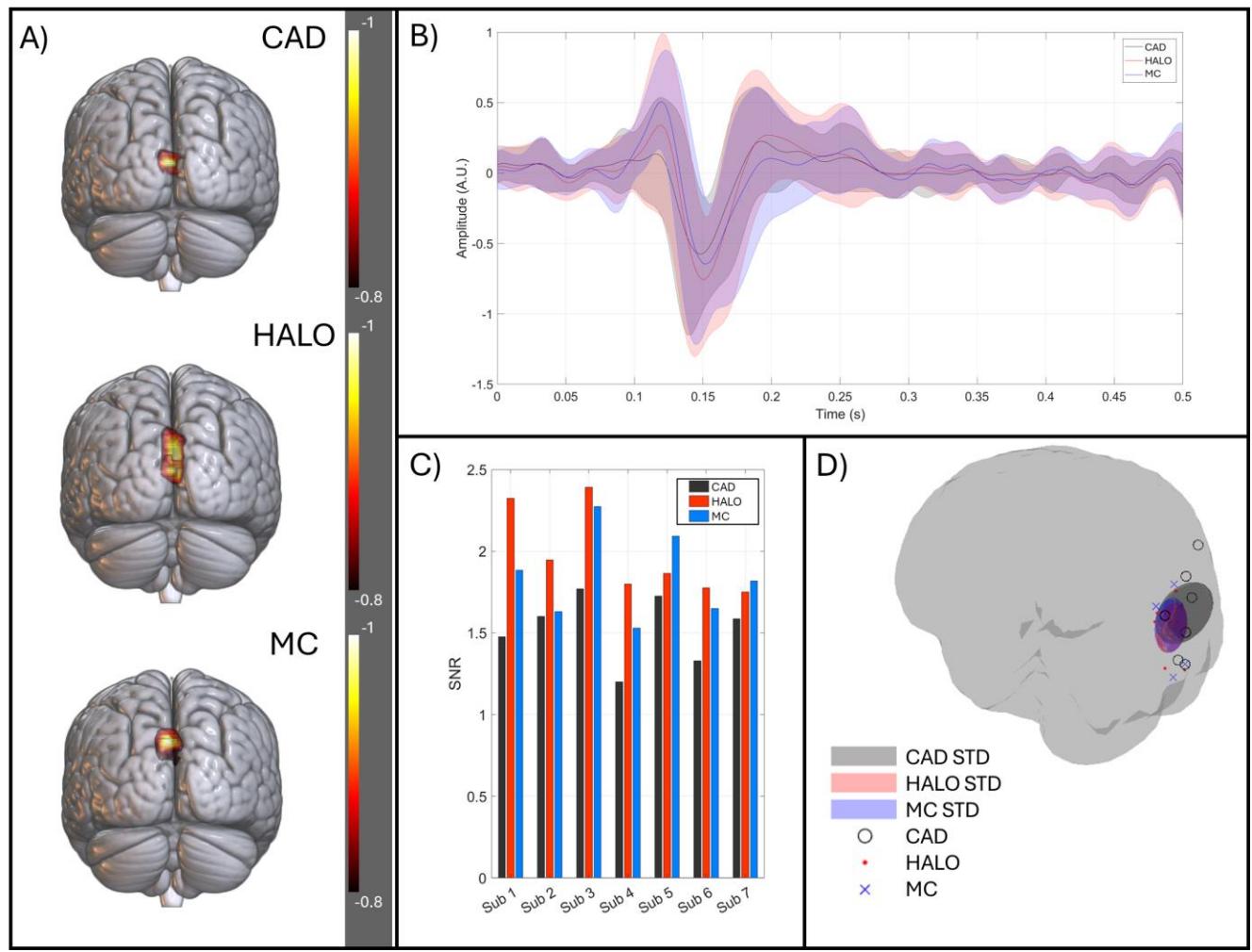

*Figure 9: Evoked responses – A) Average pseudo-Z-statistical images (normalised by their maximum absolute value) taken at t = 150 ms and thresholded at 80% of the maximum. B) the source-reconstructed evoked response, shown for each calibration method (CAD, HALO and MC in black, red and blue respectively). All data are averaged across subjects with the standard deviation shown as a shaded area. The data were reconstructed for a point at the centre of mass of the calcarine, identified by the AAL atlas. C) SNR for each subject, evaluated for CAD-, HALO-, and MC-derived calibration (in black, red and blue respectively). D) The location of the largest change in evoked response at 150 ms; peaks for each participant are shown for the 3 calibration methods. CAD shown as black circles, HALO as red dots, and MC as blue crosses. In addition, the mean position and standard deviation in each axis are shown by shaded ellipsoids.*

## 4) DISCUSSION

MEG provides a sensitive means to characterise human brain function, and has significant clinical utility, for example to identify abnormal brain activity in epilepsy and determine its spatial origin (Rampp et al., 2019). However, the accuracy of MEG-derived metrics of both normal and abnormal brain function depends critically on knowing the relative locations of sensors, their sensitive axes, and their gain. If these key calibration parameters are incorrect, systematic errors in source localisation and reduced sensitivity necessarily follow. In this paper, we demonstrate two independent techniques that use electromagnetic coils to produce known magnetic fields which can be used for OPM-MEG array calibration. We have shown that both techniques enable accurate determination of sensor location, orientation and gain; the two methods agree with each other, and both offer improvements in signal to noise ratio when characterising well-known neurophysiological processes in human MEG data.

The accuracy of calibration was assessed via measurement of the agreement between CAD-derived and calibration-derived sensor locations. This measurement was used because the CAD locations were the closest available metric to a "ground truth". We assessed sensor-to-sensor distance for all sensor pairs, and the difference between CAD- and calibration-derived values was 2.08 ± 0.07 mm and 2.6 ± 0.2 mm for HALO and MC, respectively. This notionally offers a quantifiable "error" in the sensor localisation (and consequently calibration) process. However, it is likely that this scale of variation originates not only from errors due to the calibration but also from inaccuracies in the CAD model of the helmet. Indeed, the high similarity between Figures 4B and 4C shows that the sensor-to-sensor pairings in which the HALO calibration differed most from CAD, were the same pairings for which the MC calibration differed from CAD. Further, the discrepancy between our calibration techniques (i.e., the HALO-to-MC difference of 1.97 ± 0.16 mm) was less than the discrepancy between either technique and CAD. Errors in the CAD model likely come from multiple sources including inaccuracies in 3D printing, warping of the OPM helmet over time, and inaccuracies in sensor placement in the helmet. It is likely that these issues could explain a 2-mm discrepancy. Our localisation errors on the 2-mm scale agreed well with those reported by Iivanainen et al. (3.3 mm) using a similar methodology (Iivanainen et al., 2022). Both the existing CAD and the calibrated sensor locations agree with the minimum accuracy suggested by Zetter and colleagues (Zetter et al., 2018), of <4 mm, required to be comparable to existing cryogenic MEG systems.

Unlike the sensor locations, we have no ground truth metric of sensor orientation, save that the orientation depends on the direction of modulation fields made by on-board coils, which we assume to be parallel with the axes of the outer casing of the sensor. We found that whilst the sensitive axes broadly pointed along the assumed direction, there were differences of ~17° for X-axes, ~5° for Y-axes and ~10° for Z-axes. In agreement with our sensor location results, these values were robust both across independent experiments using the same calibration procedure, and across our two (independent) calibration techniques. It is noteworthy that Zetter et al. found that channel orientation should be known to a precision of <10° for MEG reconstruction; this would be met by most (but not all) channels using the CAD determined orientations but would be met (and bettered) in full following calibration.

In the human MEG experiments, all methods allowed sources to be localised to the expected cortical regions, demonstrating that even without calibration, MEG analysis was robust. However, several measurable improvements were afforded by performing the calibration. Firstly, the maximum pseudo-T-statistics increased, implying a better source reconstruction. Secondly, the spatial spread of peak locations (measured by the ellipsoid volumes) tended to be smaller following calibration (at least for beta and evoked responses). Perhaps most importantly, the SNR of all three measured MEG responses was increased by calibration. The improvements in source space sensitivity and spatial precision result from the beamformer algorithm's exploitation of the improved match of the data to the MEG forward model after calibration. This may also explain why the improvement was larger for the beta band than it was for the gamma band. In the beta band, movement of a single digit likely evokes activity in a relatively small region of the motor cortex, and consequently the magnetic field generated looks dipolar in nature. In the gamma band, the visual stimulus was large and centrally presented, which likely evokes activity in both left and right hemispheres. This larger volume of activation is expected to generate a field that is less dipolar. If this departure from a dipole model is a larger source of error than that introduced by inaccurate calibration, then the improvement afforded by calibration would be relatively subtle. It is therefore possible (even likely) that calibration will not improve all MEG localisations.

Here, we showed that calibration improves beamformer localisation of induced and evoked effects. However, improvements should be maintained across other processing techniques. For example, the ability of the signal space separation method (SSS) (Taulu et al., 2005; Taulu and Kajola, 2005) and similar magnetic multipole modelling approaches (including homogeneous field correction (HFC) (Tierney et al., 2021) and adaptive multipole models (AMM) (Tierney et al., 2024)) to reject interference should be improved by accurate calibration due to an increased capability to discriminate spatially the measurable field patterns across the helmet (Nurminen et al., 2008). It has been shown that multi-axis sensors, especially triaxial sensors as used in the current generation of OPM-MEG devices, offer distinct advantages when applying such models (Nurminen et al., 2013; Tierney et al., 2024, 2022) (compared to sensors which only measure a single component of field) as the angle between the subspace of harmonics which span the neuronal signals and interference is dramatically increased (from ~10° in a conventional MEG system (Taulu et al., 2005) to ~60° in the OPM-MEG system presented here (Holmes et al., 2023a)). However, temporal extensions to the purely spatial models (e.g. tSSS (Taulu and Simola, 2006)) have so far been required to fully separate brain signal from background fields, implying calibration errors hinder performance (Holmes et al., 2024; Tierney et al., 2024). Future work should look to verify this improvement.

Finally, in terms of practicality each calibration process (HALO and MC) took a similar time (c.1 min) to complete. Calibration could therefore easily be performed at the beginning of a scan with relatively little disruption to the participant or experimenters. However, the repeatability of results implies that calibration need not necessarily be performed prior to every experiment. Here we showed that calibrations were stable across multiple repeats in a single day, implying one calibration per day would be sufficient. Future work may offer further insight into stability over longer periods (e.g. could calibration be performed only once per week?). However, this would only be practical if OPMs stayed in the same relative locations in a rigid helmet. In the case of a flexible (EEG-like) or adaptable OPM mounting, having a calibration procedure that can be applied every time the sensor array morphology is changed is critical. Related to this, the HALO has the advantage that it is attached to the helmet, and therefore calibration can be carried out even in the presence of large subject head movement. In contrast, the MC calibration accuracy would decrease with subject motion during the calibration. Whether such movement could be accounted for remains a topic for further study. According to the sensor localisation results, the HALO was marginally more accurate than the MC and this was generally reflected in the human MEG experiments. It is possible that the MC performed worse as the true variation of the field produced by each of the coils was not captured by the 40 x 40 x 40 $cm^3$ 10-cm resolution grid used to record the data for the MC forward model. (A finer spacing and wider volume could improve this). Furthermore, the MC are at least 1.4 m away from the sensors, meaning their spatial variation may be too slow by the time they reach the sensors to discriminate sensor locations. This said, it is noteworthy that, for the HALO, the accuracy of sensor locations drops with distance from the HALO (see appendix), perhaps as the field amplitude from the spiral coil becomes more dipolar with distance, deviating from the nearby moment (which was calibrated with a fluxgate sensor). This could be solved by mounting coils into the helmet itself, rather than on a plane riding above the helmet. For example, Adachi and colleagues embedded small coils into planar magnetoresistive sensor arrays developed for magnetocardiography (MCG) (Adachi et al., 2019). Coils for calibration could also be worn by the participant, as demonstrated by Pfeiffer et al. (Pfeiffer et al., 2020, 2018). Though we note that with such systems the proximity of the coils to the OPMs will generate focal field patterns across the sensor array. This will necessitate an improved starting guess for the optimisation algorithm than was needed in our case to avoid finding a solution in a local minimum. For our methods, the field patterns from each coil in the HALO and the MC vary similarly across the array, allowing simple navigation of the search space by the minimisation algorithm to reach the global minimum.

## 5) CONCLUSION

We have presented two methods to achieve complete OPM-MEG system calibration (measuring sensor location, orientation and gain). Both methods were based on accurately generating well-characterised magnetic fields over the sensor array. Our results show that these techniques offer an accurate means to calibrate an OPM array, and that the two methods agree strongly with each other. Further, when applied to human MEG experiments, both methods offer significantly improved SNR compared to an assumed calibration. As OPM-MEG matures, arrays become denser, channel count increases, and the technology becomes more widely used for clinical applications, the importance of calibration to obtain reliable data will increase. Technologies for calibration will therefore likely become a core part of an OPM-MEG system.

**6) ACKNOWLEDGEMENTS**

The authors would like to thank Peter D. D. Schwindt for extremely useful discussions regarding this work.

**7) FUNDING**

This work was supported by an Innovate UK biomedical catalyst grant (10037425) awarded to Cerca Magnetics Limited. We also acknowledge support from the UK Quantum Technology Hub in Sensing and Timing, funded by EPSRC (EP/T001046/1), a Medical Research Council (MRC) Mid-Range Equipment grant (MC_PC_MR/X012263/1) and an Engineering and Physical Sciences Research Council (EPSRC) Healthcare Impact Partnership Grant (EP/V047264/1). Sensor development was made possible by funding from the National Institutes of Health (R44MH110288).

**8) ETHICS**

This study was approved by the University of Nottingham Faculty of Medicine and Health Sciences Research Ethics Committee (approval number H16122016). All participants provided written informed consent.

**9) CONFLICTS OF INTEREST**

V.S. is the founding director of QuSpin, a commercial entity selling OPM magnetometers. J.O., D.B. and C.D. are employees of QuSpin. E.B. and M.J.B. are directors of Cerca Magnetics Limited, a spin-out company whose aim is to commercialise aspects of OPM-MEG technology. E.B., M.J.B., R.B., N.H. and R.H. hold founding equity in Cerca Magnetics Limited. HS and ZT are employees of Cerca Magnetics Limited.

**10) DATA AND CODE AVAILABILITY**

All data were acquired by the authors and code was custom developed in-house using MATLAB. These will be shared following a formal data sharing agreement with the authors due to subject anonymity and intellectual property.

**11) CRediT AUTHOR CONTRIBUTIONS STATEMENT**

Conceptualisation: RMH, RB, MJB, VS, NH.

Methodology: RMH, MJB, VS, NH.

Software: RMH, GRR, AT, HS, CD, JO, DB, LR, NH.

Validation: RMH, GRR, AT, HS, LR, JG, ZT.

Formal Analysis: RMH, GRR, AT, HS, CD, JO, DB, LR, JG, ZT, NH.

Investigation: RMH, GRR, AT, HS, NH.

Resources: RMH, GRR, AT, HS, CD, JO, DB, EB, NH.

Data Curation: RMH, GRR, AT, HS, CD, JO, DB, NH.

Writing – Original Draft: RMH, MJB, VS, NH.

Writing – Reviewing and Editing: All authors.

Visualisation: RMH, MJB, NH.

Supervision: EB, RB, MJB, VS, NH.

Project Administration: RMH, NH.

Funding Acquisition: EB, RB, MJB, VS.

## 12) REFERENCES

Adachi, Y., Oyama, D., Higuchi, M., Uehara, G., 2023. A Spherical Coil Array for the Calibration of Whole-Head Magnetoencephalograph Systems. IEEE Trans Instrum Meas 72, 1–10. https://doi.org/10.1109/TIM.2023.3265750

Adachi, Y., Oyama, D., Terazono, Y., Hayashi, T., Shibuya, T., Kawabata, S., 2019. Calibration of Room Temperature Magnetic Sensor Array for Biomagnetic Measurement. IEEE Trans Magn 55, 1–6. https://doi.org/10.1109/TMAG.2019.2895355

Allred, J.C., Lyman, R.N., Kornack, T.W., Romalis, M. V, 2002. High-Sensitivity Atomic Magnetometer Unaffected by Spin-Exchange Relaxation. Phys Rev Lett 89. https://doi.org/10.1103/PhysRevLett.89.130801

Altarev, I., Fierlinger, P., Lins, T., Marino, M.G., Nießen, B., Petzoldt, G., Reisner, M., Stuiber, S., Sturm, M., Taggart Singh, J., Taubenheim, B., Rohrer, H.K., Schläpfer, U., 2015. Minimizing magnetic fields for precision experiments. J Appl Phys 117. https://doi.org/10.1063/1.4922671

Baillet, S., 2017. Magnetoencephalography for brain electrophysiology and imaging. Nat Neurosci 20. https://doi.org/10.1038/nn.4504

Bentin, S., Allison, T., Puce, A., Perez, E., McCarthy, G., 1996. Electrophysiological Studies of Face Perception in Humans. J Cogn Neurosci 8. https://doi.org/10.1162/jocn.1996.8.6.551

Borna, A., Iivanainen, J., Carter, T.R., McKay, J., Taulu, S., Stephen, J., Schwindt, P.D.D., 2022. Cross-Axis projection error in optically pumped magnetometers and its implication for magnetoencephalography systems. Neuroimage 247, 118818. https://doi.org/10.1016/j.neuroimage.2021.118818

Boto, E., Holmes, N., Leggett, J., Roberts, G., Shah, V., Meyer, S.S., Muñoz, L.D., Mullinger, K.J., Tierney, T.M., Bestmann, S., Barnes, G.R., Bowtell, R., Brookes, M.J., 2018. Moving magnetoencephalography towards real-world applications with a wearable system. Nature 555. https://doi.org/10.1038/nature26147

Boto, E., Shah, V., Hill, R.M., Rhodes, N., Osborne, J., Doyle, C., Holmes, N., Rea, M., Leggett, J., Bowtell, R., Brookes, M.J., 2022. Triaxial detection of the neuromagnetic field using optically-pumped magnetometry: feasibility and application in children. Neuroimage 252, 119027. https://doi.org/10.1016/j.neuroimage.2022.119027

Brookes, M.J., Leggett, J., Rea, M., Hill, R.M., Holmes, N., Boto, E., Bowtell, R., 2022. Magnetoencephalography with optically pumped magnetometers (OPM-MEG): the next generation of functional neuroimaging. Trends Neurosci 45, 621–634. https://doi.org/10.1016/j.tins.2022.05.008

Chella, F., Zappasodi, F., Marzetti, L., Penna, S. Della, Pizzella, V., 2012. Calibration of a multichannel MEG system based on the Signal Space Separation method. Phys Med Biol 57, 4855–4870. https://doi.org/10.1088/0031-9155/57/15/4855

Cohen-Tannoudji, C., Dupont-Roc, J., Haroche, S., Laloë, F., 1970. Diverses résonances de croisement de niveaux sur des atomes pompés optiquement en champ nul. I. Théorie. Revue de Physique Appliquée. https://doi.org/10.1051/rphysap:019700050109500

Corvilain, P., Wens, V., Bourguignon, M., Capparini, C., Fourdin, L., Ferez, M., Ferez, M., Feys, O., De Tiège, X., Bertels, J., 2024. Extending the applicability of optically pumped magnetoencephalography toward early human life. bioRxiv.

Feys, O., Corvilain, P., Aeby, A., Sculier, C., Holmes, N., Brookes, M., Goldman, S., Wens, V., De Tiège, X., 2022. On-Scalp Optically Pumped Magnetometers versus Cryogenic Magnetoencephalography for Diagnostic Evaluation of Epilepsy in School-aged Children. Radiology 304, 429–434. https://doi.org/10.1148/radiol.212453

Halgren, E., Raij, T., Marinkovic, K., Jousmäki, V., Hari, R., 2000. Cognitive Response Profile of the Human Fusiform Face Area as Determined by MEG. Cerebral Cortex 10. https://doi.org/10.1093/cercor/10.1.69

Hämäläinen, M., Hari, R., Ilmoniemi, R.J., Knuutila, J., Lounasmaa, O. V, 1993. Magnetoencephalography—theory, instrumentation, and applications to noninvasive studies of the working human brain. Rev Mod Phys 65. https://doi.org/10.1103/RevModPhys.65.413

Hammond, P., 1960. Electric and magnetic images. Proceedings of the IEE Part C: Monographs 107, 306. https://doi.org/10.1049/pi-c.1960.0047

Happer, W., 1972. Optical pumping. Rev Mod Phys. https://doi.org/10.1103/RevModPhys.44.169

Hill, R.M., Boto, E., Holmes, N., Hartley, C., Seedat, Z.A., Leggett, J., Roberts, G., Shah, V., Tierney, T.M., Woolrich, M.W., Stagg, C.J., Barnes, G.R., Bowtell, R., Slater, R., Brookes, M.J., 2019. A tool for functional brain imaging with lifespan compliance. Nat Commun 10. https://doi.org/10.1038/s41467-019-12486-x

Hill, R.M., Boto, E., Rea, M., Holmes, N., Leggett, J., Coles, L.A., Papastavrou, M., Everton, S.K., Hunt, B.A.E., Sims, D., Osborne, J., Shah, V., Bowtell, R., Brookes, M.J., 2020. Multi-channel whole-head OPM-MEG: Helmet design and a comparison with a conventional system. Neuroimage 219. https://doi.org/10.1016/j.neuroimage.2020.116995

Hill, Ryan M, Devasagayam, J., Holmes, N., Boto, E., Shah, V., Osborne, J., Safar, K., Worcester, F., Mariani, C., Dawson, E., Woolger, D., Bowtell, R., Taylor, M.J., Brookes, M.J., 2022. Using OPM-MEG in contrasting magnetic environments. Neuroimage 253, 119084. https://doi.org/10.1016/j.neuroimage.2022.119084

Hill, R.M., Schofield, H., Boto, E., Rier, L., Osborne, J., Doyle, C., Worcester, F., Hayward, T., Holmes, N., Bowtell, R., Shah, V., Brookes, M.J., 2024. Optimising the Sensitivity of Optically-Pumped Magnetometer Magnetoencephalography to Gamma Band Electrophysiological Activity. Imaging Neuroscience.

Holmes, N., Bowtell, R., Brookes, M.J., Taulu, S., 2023a. An Iterative Implementation of the Signal Space Separation Method for Magnetoencephalography Systems with Low Channel Counts. Sensors 23. https://doi.org/10.3390/s23146537

Holmes, N., Leggett, J., Hill, R.M., Rier, L., Boto, E., Schofield, H., Hayward, T., Dawson, E., Woolger, D., Shah, V., Taulu, S., Brookes, M.J., Bowtell, R., 2024. Wearable magnetoencephalography in a lightly shielded environment. IEEE Trans Biomed Eng 1–10. https://doi.org/10.1109/TBME.2024.3465654

Holmes, N., Rea, M., Chalmers, J., Leggett, J., Edwards, L.J., Nell, P., Pink, S., Patel, P., Wood, J., Murby, N., Woolger, D., Dawson, E., Mariani, C., Tierney, T.M., Mellor, S., O'Neill, G.C., Boto, E., Hill, R.M., Shah, V., Osborne, J., Pardington, R., Fierlinger, P., Barnes, G.R., Glover, P., Brookes, M.J., Bowtell, R., 2022. A lightweight magnetically shielded room with active shielding. Sci Rep 12, 13561. https://doi.org/10.1038/s41598-022-17346-1

Holmes, N., Rea, M., Hill, R.M., Leggett, J., Edwards, L.J., Hobson, P.J., Boto, E., Tierney, T.M., Rier, L., Rivero, G.R., Shah, V., Osborne, J., Fromhold, T.M., Glover, P., Brookes, M.J., Bowtell, R., 2023b. Enabling ambulatory movement in wearable magnetoencephalography with matrix coil active magnetic shielding. Neuroimage 274, 120157. https://doi.org/10.1016/j.neuroimage.2023.120157

Hoogenboom, N., Schoffelen, J.-M., Oostenveld, R., Parkes, L.M., Fries, P., 2006. Localizing human visual gamma-band activity in frequency, time and space. Neuroimage 29. https://doi.org/10.1016/j.neuroimage.2005.08.043

Iivanainen, J., Borna, A., Zetter, R., Carter, T.R., Stephen, J.M., McKay, J., Parkkonen, L., Taulu, S., Schwindt, P.D.D., 2022. Calibration and Localization of Optically Pumped Magnetometers Using Electromagnetic Coils. Sensors 22. https://doi.org/10.3390/s22083059

Iivanainen, J., Stenroos, M., Parkkonen, L., 2017. Measuring MEG closer to the brain: Performance of on-scalp sensor arrays. Neuroimage 147. https://doi.org/10.1016/j.neuroimage.2016.12.048

Jenkinson, M., Bannister, P., Brady M, Smith, S., 2002. Improved Optimization for the Robust and Accurate Linear Registration and Motion Correction of Brain Images. Neuroimage 17, 825–841. https://doi.org/10.1016/S1053-8119(02)91132-8

Jenkinson, M., Smith, S., 2001. A global optimisation method for robust affine registration of brain images. Med Image Anal 5, 143–156. https://doi.org/10.1016/S1361-8415(01)00036-6

Li, H., Zhang, S.-L., Zhang, C.-X., Kong, X.-Y., Xie, X.-M., 2016. An efficient calibration method for SQUID measurement system using three orthogonal Helmholtz coils. Chinese Physics B 25, 068501. https://doi.org/10.1088/1674-1056/25/6/068501

Mellor, S., Tierney, T.M., Seymour, R.A., Timms, R.C., O'Neill, G.C., Alexander, N., Spedden, M.E., Payne, H., Barnes, G.R., 2023. Real-time, model-based magnetic field correction for moving, wearable MEG. Neuroimage 278, 120252. https://doi.org/10.1016/j.neuroimage.2023.120252

Mellor, S.J., Tierney, T., O'Neill, G., Alexander, N., Seymour, R., Holmes, N., Lopez, J.D., Hill, R., Boto, E., Rea, M., Roberts, G., Leggett, J., Bowtell, R., Brookes, M.J., Maguire, E., Walker, M., Barnes, G., 2021. Magnetic Field Mapping and Correction for Moving OP-MEG. IEEE Trans Biomed Eng 9294. https://doi.org/10.1109/TBME.2021.3100770

Nolte, G., 2003. The magnetic lead field theorem in the quasi-static approximation and its use for magnetoencephalography forward calculation in realistic volume conductors. Phys Med Biol 48. https://doi.org/10.1088/0031-9155/48/22/002

Nurminen, J., Taulu, S., Nenonen, J., Helle, L., Simola, J., Ahonen, A., 2013. Improving MEG performance with additional tangential sensors. IEEE Trans Biomed Eng 60, 2559–2566. https://doi.org/10.1109/TBME.2013.2260541

Nurminen, J., Taulu, S., Okada, Y., 2008. Effects of sensor calibration, balancing and parametrization on the signal space separation method. Phys Med Biol 53, 1975–1987. https://doi.org/10.1088/0031-9155/53/7/012

Oyama, D., Adachi, Y., Higuchi, M., Uehara, G., 2022. Calibration of a Coil Array Geometry Using an X-Ray Computed Tomography. IEEE Trans Magn 58, 1–5. https://doi.org/10.1109/TMAG.2021.3080673

Pfeiffer, C., Andersen, L.M., Lundqvist, D., Hämäläinen, M., Schneiderman, J.F., Oostenveld, R., 2018. Localizing on-scalp MEG sensors using an array of magnetic dipole coils. PLoS One 13, e0191111. https://doi.org/10.1371/journal.pone.0191111

Pfeiffer, C., Ruffieux, S., Andersen, L.M., Kalabukhov, A., Winkler, D., Oostenveld, R., Lundqvist, D., Schneiderman, J.F., 2020. On-scalp MEG sensor localization using magnetic dipole-like coils: A method for highly accurate co-registration. Neuroimage 212, 116686. https://doi.org/10.1016/j.neuroimage.2020.116686

Pfurtscheller, G., Lopes da Silva, F.H., 1999. Event-related EEG/MEG synchronization and desynchronization: basic principles. Clinical Neurophysiology 110, 1842–1857. https://doi.org/10.1016/S1388-2457(99)00141-8

Rampp, S., Stefan, H., Wu, X., Kaltenhäuser, M., Maess, B., Schmitt, F.C., Wolters, C.H., Hamer, H., Kasper, B.S., Schwab, S., Doerfler, A., Blümcke, I., Rössler, K., Buchfelder, M., 2019. Magnetoencephalography for epileptic focus localization in a series of 1000 cases. Brain 142, 3059–3071. https://doi.org/10.1093/brain/awz231

Rea, M., Holmes, N., Hill, R.M., Boto, E., Leggett, J., Edwards, L.J., Woolger, D., Dawson, E., Shah, V., Osborne, J., Bowtell, R., Brookes, M.J., 2021. Precision magnetic field modelling and control for wearable magnetoencephalography. Neuroimage 241. https://doi.org/10.1016/j.neuroimage.2021.118401

Rier, L., Rhodes, N., Pakenham, D., Boto, E., Holmes, N., Hill, R.M., Reina Rivero, G., Shah, V., Doyle, C., Osborne, J., Bowtell, R., Taylor, M.J., Brookes, M.J., 2024. The neurodevelopmental trajectory of beta band oscillations: an OPM-MEG study. Elife.

Robinson, S., Vrba, J., 1998. Functional Neuroimaging by synthetic Aperture Magnetometry. Recent Advances in Biomagnetism 302–305.

Roshen, W.A., 1990. Effect of Finite Thickness of Magnetic Substrate on Planar Inductors. IEEE Trans Magn 26, 270–275. https://doi.org/10.1109/20.50553

Schofield, H., Boto, E., Shah, V., Hill, R.M., Osborne, J., Rea, M., Doyle, C., Holmes, N., Bowtell, R., Woolger, D., Brookes, M.J., 2023. Quantum enabled functional neuroimaging: the why and how of magnetoencephalography using optically pumped magnetometers. Contemp Phys 63, 161–179. https://doi.org/10.1080/00107514.2023.2182950

Schofield, H., Hill, R.M., Feys, O., Holmes, N., Osborne, J., Doyle, C., Bobela, D., Corvilain, P., Wens, V., Rier, L., Bowtell, R., Ferez, M., Mullinger, K.J., Coleman, S., Rhodes, N., Rea, M., Tanner, Z., Boto, E., de Tiège, X., Shah, V., Brookes, M.J., 2024. A novel, robust, and portable platform for magnetoencephalography using optically-pumped magnetometers. Imaging Neuroscience 2, 1–22. https://doi.org/10.1162/imag_a_00283

Schwindt, P.D.D., Knappe, S., Shah, V., Hollberg, L., Kitching, J., Liew, L.A., Moreland, J., 2004. Chip-scale atomic magnetometer. Appl Phys Lett. https://doi.org/10.1063/1.1839274

Schwindt, P.D.D., Lindseth, B., Knappe, S., Shah, V., Kitching, J., Liew, L.-A., 2007. Chip-scale atomic magnetometer with improved sensitivity by use of the Mx technique. Appl Phys Lett 90. https://doi.org/10.1063/1.2709532

Shah, V., Doyle, C., Osborne, J., 2020. Zero field parametric resonance magnetometer with triaxial sensitivity.

Shah, V., Knappe, S., Schwindt, P.D.D., Kitching, J., 2007. Subpicotesla atomic magnetometry with a microfabricated vapour cell. Nat Photonics 1, 649–652. https://doi.org/10.1038/nphoton.2007.201

Shah, V.K., Wakai, R.T., 2013. A compact, high performance atomic magnetometer for biomedical applications. Phys Med Biol 58, 8153–8161. https://doi.org/10.1088/0031-9155/58/22/8153

Taulu, S., Kajola, M., 2005. Presentation of electromagnetic multichannel data: The signal space separation method. J Appl Phys 97. https://doi.org/10.1063/1.1935742

Taulu, S., Simola, J., 2006. Spatiotemporal signal space separation method for rejecting nearby interference in MEG measurements. Phys Med Biol 51, 1759–1768. https://doi.org/10.1088/0031-9155/51/7/008

Taulu, S., Simola, J., Kajola, M., 2005. Applications of the signal space separation method. IEEE Transactions on Signal Processing 53, 3359–3372. https://doi.org/10.1109/TSP.2005.853302

Taylor, M.J., George, N., Ducorps, A., 2001. Magnetoencephalographic evidence of early processing of direction of gaze in humans. Neurosci Lett 316. https://doi.org/10.1016/S0304-3940(01)02378-3

Tierney, T.M., Alexander, N., Mellor, S., Holmes, N., Seymour, R., O'Neill, G.C., Maguire, E.A., Barnes, G.R., 2021. Modelling optically pumped magnetometer interference in MEG as a spatially homogeneous magnetic field. Neuroimage 244. https://doi.org/10.1016/j.neuroimage.2021.118484

Tierney, T.M., Barnes, G.R., Seedat, Z., Pier, K.S., Mellor, S., 2024. Adaptive multipole models of optically pumped magnetometer data 1–16. https://doi.org/10.1002/hbm.26596

Tierney, T.M., Holmes, N., Mellor, S., López, J.D., Roberts, G., Hill, R.M., Boto, E., Leggett, J., Shah, V., Brookes, M.J., Bowtell, R., Barnes, G.R., 2019. Optically pumped magnetometers: From quantum origins to multi-channel magnetoencephalography. Neuroimage 199, 598–608. https://doi.org/10.1016/j.neuroimage.2019.05.063

Tierney, T.M., Mellor, S., O'Neill, G.C., Timms, R.C., Barnes, G.R., 2022. Spherical harmonic based noise rejection and neuronal sampling with multi-axis OPMs. Neuroimage 258, 119338. https://doi.org/10.1016/j.neuroimage.2022.119338

Vrba, J., Robinson, S.E., 2001. Signal Processing in Magnetoencephalography. Methods 25, 249–271. https://doi.org/10.1006/meth.2001.1238

Zetter, R., Iivanainen, J., Stenroos, M., Parkkonen, L., 2018. Requirements for Coregistration Accuracy in On-Scalp MEG. Brain Topogr. https://doi.org/10.1007/s10548-018-0656-5

## 13) APPENDIX: Sensor localisation accuracy with distance from the HALO

Figure A1a shows Euclidean distance between the HALO-derived and the CAD-derived sensor locations, following application of the ICP algorithm (y-axis), plotted against the distance between the sensor and the centre of the HALO (x-axis). Figure A1b shows the equivalent data for the MC-derived sensor locations. In both cases, data from all experimental runs are overlaid. Both plots also show a line of best-fit, with an R-squared value of 0.18 for the HALO and 0.0142 for the MC. The relationship in Figure A1a suggests that sensors distal to the HALO tended to localise more poorly than proximal sensors. As would be expected this relationship is absent for the MC-derived sensor locations.

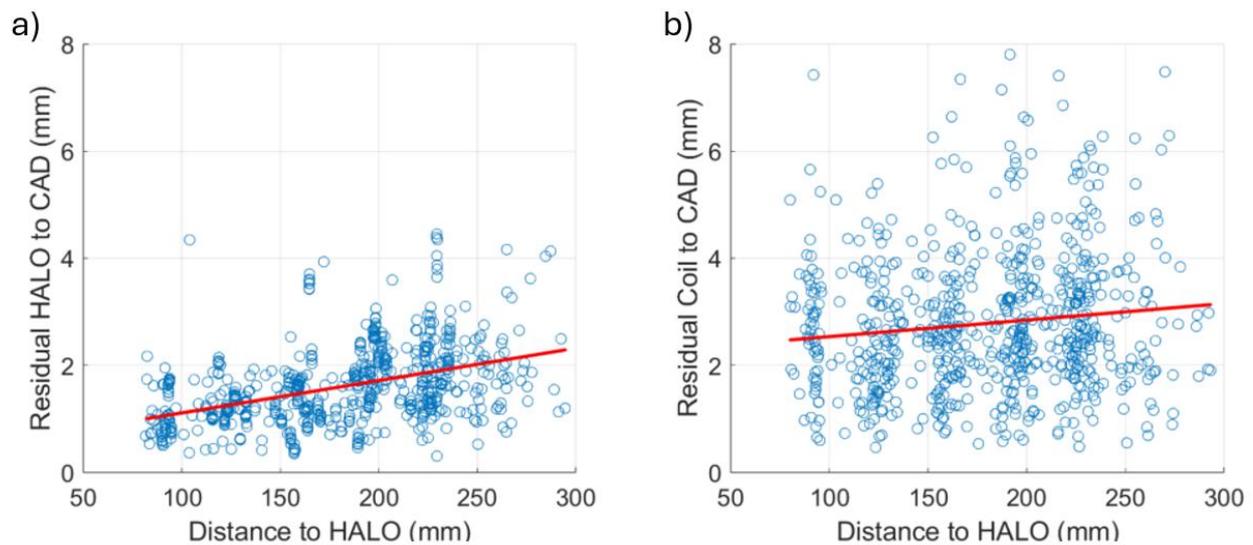

*Figure A1: a) Absolute distances between the CAD and HALO sensor positions following application of the ICP algorithm, plotted against distance from the sensor to the centre of the HALO. b) equivalent data for the MC. Note a reduction in agreement with distance for HALO calibration.*